\documentclass[aps,pre,twocolumn]{revtex4-1}
\usepackage{bm,epsfig,graphics,amssymb,amsmath,subeqnarray,setspace,graphicx,amsthm,epstopdf,subfigure,color}
\usepackage{epstopdf, epsf,psfrag, graphicx,color}

\def\b{\mathbf}\def\ep{\varepsilon}

\def\tcb{\textcolor{black}}

\begin{document}
\title{{Geometric capture and escape of a microswimmer colliding with an obstacle}}
\author{Saverio E. Spagnolie$^{1}$, Gregorio R. Moreno-Flores$^{1,2}$, Denis Bartolo$^{3}$ and Eric Lauga$^{4}$}
\affiliation{$^{1}$Department of Mathematics, University of Wisconsin-Madison, 480 Lincoln Dr., Madison, WI 53706}
\affiliation{$^{2}$Departamento de Matemáticas, Pontificia Universidad Catolica de Chile}
\affiliation{$^{3}$ {Laboratoire de Physique ENS de Lyon, UniversitŽ de Lyon, 46, all\'ee d'Italie 69007 Lyon, France}}
\affiliation{$^{4}$Department of Applied Mathematics and Theoretical Physics, University of Cambridge, Wilberforce Road, Cambridge CB3 0WA, United Kingdom}
\date{\today}

\begin{abstract}
Motivated by recent experiments, we consider the hydrodynamic capture of a microswimmer near a stationary spherical obstacle. Simulations of model equations show that a swimmer approaching a small spherical colloid is simply scattered. In contrast, when the colloid is larger than a critical size it acts as a passive trap:  the swimmer is hydrodynamically captured along closed trajectories and endlessly orbits around the colloidal sphere. In order to gain physical insight into this hydrodynamic scattering problem, we address it analytically. We provide expressions for the critical trapping radius, the depth of the ``basin of attraction,'' and the scattering angle, which show excellent agreement with our numerical findings. We also demonstrate and rationalize the strong impact of swimming-flow symmetries on the trapping efficiency.  Finally, we give the swimmer an opportunity to escape the colloidal traps by considering the effects of Brownian, or active, diffusion. We show that in some cases the trapping time is governed by an Ornstein-Uhlenbeck process, which results in a trapping time distribution that is well-approximated as inverse-Gaussian. The predictions again compare very favorably with the numerical simulations. We envision applications of the theory to bioremediation, microorganism sorting techniques, and the study of bacterial populations in heterogeneous or porous environments.
\end{abstract}
\maketitle

\section{Introduction}

Microorganisms and other self-propelling bodies in viscous fluids are known to traverse complex trajectories in the presence of boundaries. One basic interaction with a plane wall, observed in experiments with {\it Escherichia coli} bacteria and spermatozoa, is that the cells may accumulate near the surface due to a combination of  hydrodynamic and steric effects \cite{Rothschild63,fm95,btbl08,sgbk09,sb09,ddcgg11,sl12}. Another effect, associated with the rotation of helical flagella and a counter-rotation of the cell body in {\it E. coli}, is that flagellated bacteria swim in large circles when they are near a solid boundary \cite{ldlws06}, and in circles of opposite handedness near a free surface \cite{dldaai11}. The orientations of swimming bodies, even those hydrodynamically bound to the surface, are non-trivial and depend on the geometry of the swimmer and its mechanism of propulsion \cite{gnbnm05,sgs10,giy10,znm09,houg09,co10,lp10,Crowdy11,sl12}.

The attraction and trapping of microorganisms near surfaces may lead to the development of biofilms \cite{vllnz90,otkk00}, and possible infection of medically implanted surfaces \cite{hdf92}. Other biophysical properties may also be important; for example, {\it Chlamydomonas} algae cells scatter from a flat wall due to contact between its flagella and the surface, so that the interaction is highly dependent on the body and flagellar lengths and geometries \cite{kdpg13}, and the tumbling of {\it E. coli} is suppressed near surfaces due to increased hydrodynamic resistance \cite{mbss14}. From a bioengineering perspective, sorting and rectification devices have also been constructed at the microscale which exploit the interactions of microorganisms and asymmetric surfaces (including funnels and gears) \cite{gkca07,wrnr08,tc09,dladariscmdadf10,bjmvdvscm13,wlst15}. In some cases, steric collisions or near-field lubrication forces may dominate long-range hydrodynamic effects \cite{ddcgg11,wwdkg13,lwg14}. 

\begin{figure}[htbp]
\begin{center}
\includegraphics[width=.3\textwidth]{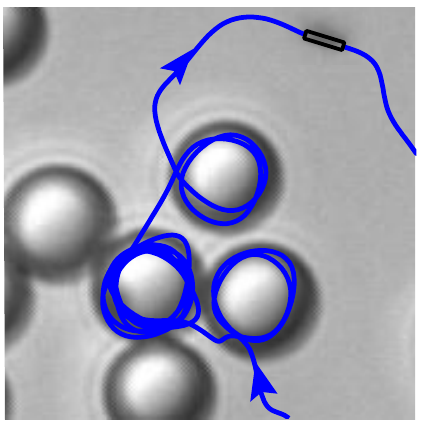}
\caption{Snapshots from the experiments of Takagi et al.~\cite{tbzm13}, reproduced with permission. The swimming trajectory of a self-propelled body in a colloid-filled bath includes a billiard-like motion between colloids, intermittent periods of entrapped and orbiting states, and randomized escape behavior.}
\label{Figure1}
\end{center}
\end{figure}

Naturally, interactions with geometrical boundaries is not specific to living organisms, and also applies to the synthetic self-propelled colloids  that have been extensively studied over the last five years~\cite{pkossacmlc04, fbamo05, rk07, gf09,pgwl11,dbrfsb05,Wang09,tbzm13}. A recent experiment by Takagi {\it et al.}~\cite{tpbsz13} {showed that a self-propelled synthetic swimmer in a field of passive colloidal beads displays its own complex trajectory.} The path includes a billiard-like motion between colloids, intermittent periods of entrapped, orbiting states near single colloids, and randomized escape behavior (see Fig.~\ref{Figure1}). Takagi {\it et al.}~\cite{tpbsz13} argued that short-range hydrodynamic interactions and steric effects were sufficient to understand their experimental results. Brown {\it et al.} explored an extension of these dynamics to swimming through a ``colloidal crystal,'' where a synthetic swimmer hops from colloid to colloid with a trapping time that depends on fuel concentration, whereas {\it E. coli} trajectories are rectified into long, straight runs \cite{bvdvslp15}.

In this article, we set out to understand quantitatively the hydrodynamic scattering of a swimming body by a stationary spherical obstacle. We develop a semi-analytical model to describe the trajectory of a model swimmer based on far-field hydrodynamic interactions {and hard-core repulsion}. Using numerical simulations  {of this minimal model, we demonstrate that:} (i) the swimmer can be hydrodynamically trapped by colloids above a critical size, (ii) sub-critical interactions involve only short residence times on the surface, and (iii) that model ``puller'' swimmers may be trapped by much smaller colloids than are necessary to trap ``pusher'' swimmers. The critical colloid size for the entrapment of pusher particles is found to scale quadratically with the inverse of the swimmer dipole strength, and for puller particles with only the inverse of the dipole strength. The residence time for sub-critical interactions is also considered, as is the size of the ``basin of attraction'' around the colloid below which a swimmer can be drawn into the surface. A scaling law for the basin radius is deduced, resulting in a {master}curve onto which all of the numerically simulated values collapse. A semi-analytical expression is also provided for the total scattering angle in the case of sub-critical colloid size. Finally, with the introduction of Brownian fluctuations, swimmers trapped in the deterministic setting are shown to escape randomly. The distribution of trapping times are analyzed for a range of colloid sizes, swimmer types, and diffusion constants. In some cases the trapping time is governed by an Ornstein-Uhlenbeck process, which results in trapping time distributions that are well-approximated as inverse-Gaussian. The predictions are again found to match the numerical simulations closely.

The paper is organized as follows. In \S\ref{sec: equations} the mathematical model is presented. Analytical formulae for swimming velocities are developed using the image singularity system of Oseen and the application of Fax\'en's Law. The resulting swimming trajectories are described in \S\ref{sec: entrapment and scattering}, where we obtain a criterion for deterministic hydrodynamic capture. In addition, the scattering dynamics is derived for near-obstacle interactions, the basin of attraction is shown to collapse to a power-law, and trapping of puller-type swimmers is shown to be possible using a much smaller colloid. In \S\ref{sec: fluct} we consider the effects of translational and rotational fluctuations, which have distinct consequences on entrapment, escape, and the statistics of swimming in random media. The trapping time distribution is explored for varying dipole strength, colloid size, and diffusion constant. We conclude with a discussion in \S\ref{sec:conc}.

\section{Mathematical model}\label{sec: equations}

We begin by describing a mathematical model for the dynamics of self-propulsion near a stationary spherical obstacle. In an unbounded fluid the body is assumed to swim unhindered at a speed $U$ along a director $\b{\hat{e}}$, but it can deviate from its straight path in the presence of a background flow $\b{u}$. For mathematical convenience, the swimmer body is assumed to take the shape of an ellipsoid with semi-major axis length $a$ and aspect ratio $\gamma$. Scaling velocities upon $U$ and lengths upon $a$, the position $\b{x}_0(t)$ and orientation $\b{\hat{e}}(t)$ of the swimmer are provided by Fax\'en's Law \cite{kk91},
\begin{gather}
\frac{d\b{x}_0}{dt}=\b{\hat{e}}+\b{\tilde{u}},\ \ \ 
\frac{d\b{\hat{e}}}{dt}=\bm{\tilde{\Omega}}\times \b{\hat{e}},\label{Eqs: Traj}
\end{gather}
where $\b{\tilde{u}}$ and $\bm{\tilde{\Omega}}$ are the hydrodynamic contributions to the dynamics which are zero in an unbounded quiescent fluid.

\begin{figure}[t]
\begin{center}
\includegraphics[width=.46\textwidth]{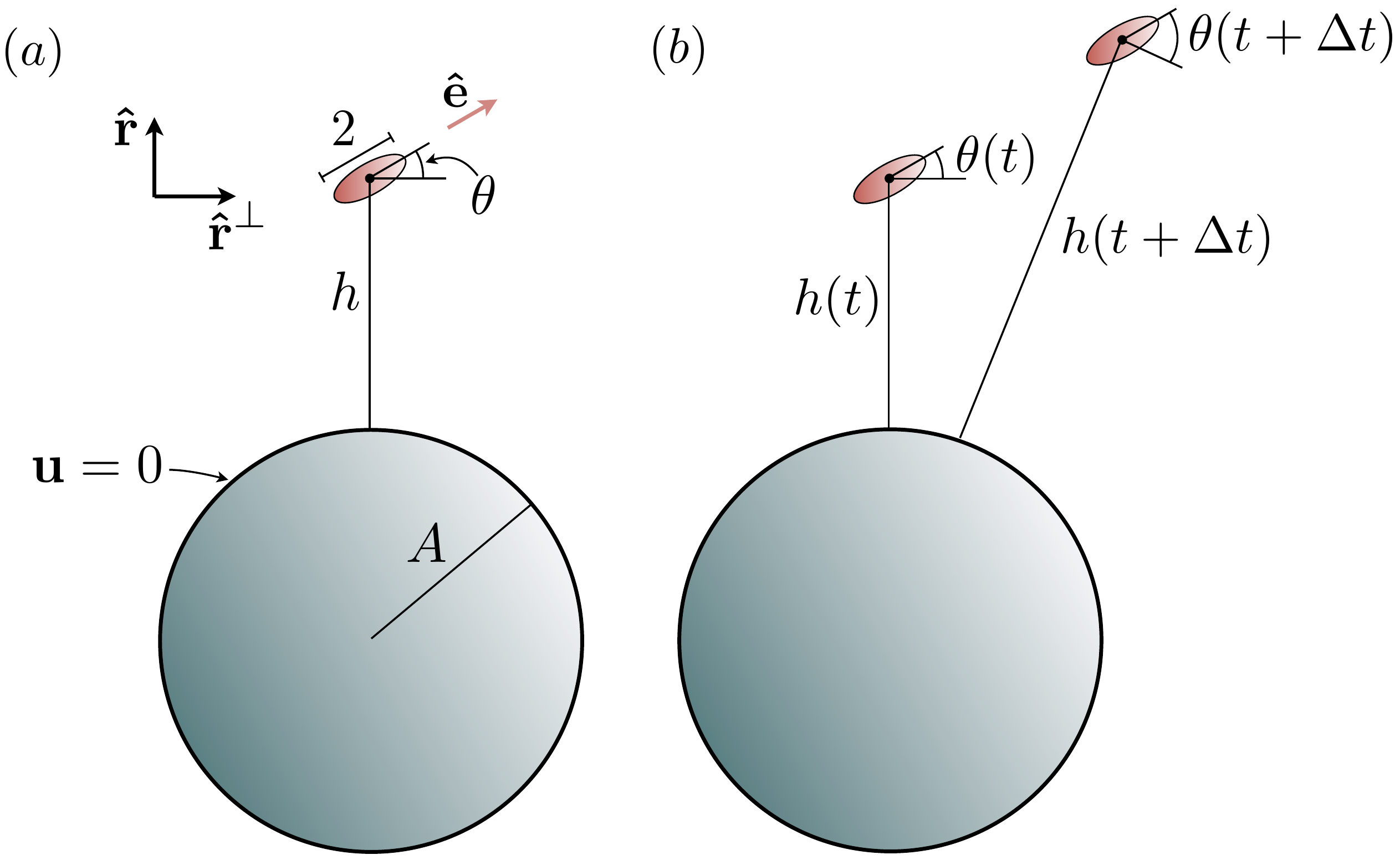}
\caption{(a) Illustration of the colloid/swimmer system. A swimming body of dimensionless length 2 swims in a direction $\b{\hat{e}}$. Its centroid lies a distance $h$ away from the surface of a spherical colloid of radius $A$. The angle between the director $\b{\hat{e}}$ and the line perpendicular to the line of centers is denoted by $\theta$. (b) The distance $h$ and relative angle $\theta$ change even when the body swims straight due to the geometry.}
\label{Colloid1}
\end{center}
\end{figure}

Consider the introduction of a single spherical colloid of dimensionless radius $A$ placed at the origin. The setup is illustrated in Fig.~\ref{Colloid1}a. The unit vectors $\b{\hat{r}}$ and $\b{\hat{r}^\perp}$ are defined at each moment in time relative to the line joining the centers of the swimmer and sphere. The angle between the swimming director $\b{\hat{e}}$ and the line perpendicular to the line of centers is denoted by $\theta$, and the centroid of the swimmer is located a distance $h$ from the colloid surface. In addition to the hydrodynamic impact on the trajectory, the distance and angle of the swimmer relative to the sphere also changes in time due simply to geometry, as illustrated in Fig.~\ref{Colloid1}b. Combining the hydrodynamic and geometric contributions to the swimming dynamics, the translational and angular swimming velocities in terms of $h$ and $\theta$ are given by
\begin{align}
\frac{dh}{dt}&=\sin(\theta)+\b{\hat{r}}\cdot \b{\tilde{u}},\label{hdot_general}\\
\frac{d\theta}{dt}&=\frac{1}{A+h}\left(\cos(\theta)+\b{\hat{r}^\perp}\cdot \b{\tilde{u}}\right)+(\b{\hat{r}}^\perp\times\b{\hat{r}})\cdot\bm{\tilde{\Omega}}.\label{tdot_general}
\end{align}
When the swimmer makes contact with the surface, we assume a simple rigid-body interaction. Specifically, when geometrical contact with the surface occurs, $\theta$ is still allowed to vary according to Eq.~\eqref{hdot_general}, but $h$ varies only if $\dot{h}>0$, so that the swimmer cannot penetrate the colloid. When the swimmer is in contact with the wall we therefore write
\begin{gather}\label{hdot:discontinuous}
\frac{dh}{dt}=\max\{\sin(\theta)+\b{\hat{r}}\cdot \b{\tilde{u}},0\}.
\end{gather}
\tcb{This is equivalent to the swimmer experiencing a hard wall repulsion (Heaviside potential) with no torque.}

\subsection{Far-field hydrodynamics}
Thus far we have not assumed anything about the ambient flow field local to the swimming body, or about the flow field generated by the {swimming motion}. 
{Let us first summarize the approach that we take in this paper in order to model the interplay between the swimmer propulsion and the fluid flow}. The flow field generated by the swimming {motion} is approximated by its leading order approximation far from the body. {This simplified flow} takes the form of a singular solution to the underlying Stokes equations of viscous fluids \cite{ss08,lp09,sl12,ss14}. Images of the fundamental singularity solutions to the Stokes equations have been used to derive flows in the upper-half plane with no-slip boundary conditions \cite{bc74,Blake71}. Those flow fields, along with an application of Fax\'en's Law, result in a description of the trajectory of a self-propelled body near a wall \cite{btbl08,ddcgg11,sl12} or a stress-free surface \cite{sl12}. A similar technique may be used to find the flow generated by a point force external to a sphere with a no-slip boundary condition, as derived by Oseen \cite{Oseen27}, and it is used here to derive the hydrodynamic effect of the colloid on the swimming body. We now describe these steps for the present case in greater detail.

Although the fluid flow near a swimming organism is complex and depends on both the swimmer geometry and the propulsive mechanism, the flow far from the body may be represented as a multipole expansion of the velocity field so produced. The flow-field far from a neutrally buoyant self-propelled body at leading order is given by 
\begin{gather}
\b{u(x)}=\alpha\, \b{S_D}(\b{x-x}_0;\b{\hat{e}})+O\left(|\b{x-x}_0|^3\right),
\end{gather}
where
\begin{gather}
\b{S_D}(\b{x,\b{\hat{e}}})=\frac{\b{x}}{|\b{x}|^3}\left(\frac{3(\b{\b{\hat{e}}\cdot x})^2}{|\b{x}|^2}-1\right)
\end{gather}
is a symmetric force dipole \cite{lp09}. The value of the coefficient $\alpha$ may be measured for a given microorganism. Recent experimental measurement of the flow produced by a swimming {\it E. coli} cell was performed by Drescher et al.~\cite{ddcgg11}, for which $\alpha$ was approximately $\alpha=0.6$. Swimmers with $\alpha>0$ are known as {\it pushers}, and those with $\alpha<0$ are known as {\it pullers} \cite{ss07}. We henceforth focus our attention on values of $\alpha$ on this scale which is also relevant to synthetic microswimmers.

\subsection{Image singularity system and method of reflections}

\begin{figure*}[t]
\begin{center}
\includegraphics[width=.9\textwidth]{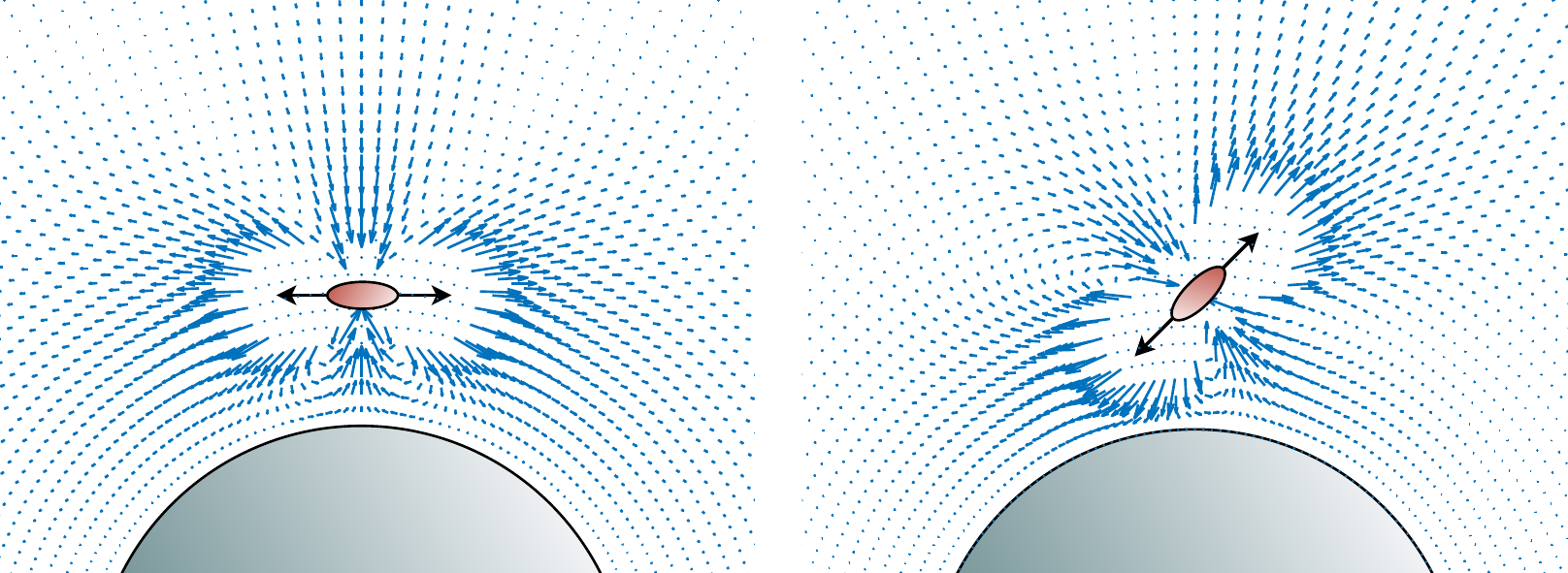}
\caption{The flow fields due to a pusher ($\alpha>0$) near a sphere, with $\theta=0$ (left) and $\theta=\pi/4$ (right). The flow field for $\theta=0$ case suggests a hydrodynamic attraction to the colloid, while the $\theta=\pi/4$ case suggests an extra hydrodynamic repulsion from the colloid. A puller  ($\alpha<0$) generates the identical  flow field but with velocity  signs reversed. The flow field is singular at the swimmer center; the velocity near the swimmer is not shown here for the sake of clarity.}
\label{FlowFields}
\end{center}
\end{figure*}

We denote the singular solutions to the Stokes equations placed internal to the spherical body, selected so as to cancel the fluid velocity on the surface $|\b{x}|=A$, by $\b{u}^*(\b{x})=\b{S^*_D}(\b{x}-\b{x}_0^*,\b{\hat{e}})$, where $\b{x}_0^*=(A^2/|\b{x}_0|^2)\b{x}_0$ is the image point of the swimming body inside of the sphere (details are given in Appendix A). By introducing the image system, the fluid flow given by
\begin{gather}
\b{u}(\b{x})=\alpha\left[\b{S_D}(\b{x-x_0;\hat{e}})+\b{S^*_D}(\b{x-x_0^*;\hat{e}})\right],
\end{gather}
is such that $\b{u}=0$ on the surface of the colloid, as shown in Fig.~\ref{FlowFields} for $\theta=0$ and $\theta=\pi/4$. The total flow no longer satisfy the appropriate boundary conditions  on the surface of the swimming body. Instead, there results a net force and torque on the swimmer associated with the image flow, which when balanced with translational and rotational drag return the leading-order hydrodynamic effect of the colloid  on the swimming trajectory. 

Returning to Eq.~\eqref{Eqs: Traj}, Fax\'en's Law for an ellipsoidal particle results in the expressions
\begin{gather}
\b{\tilde{u}}=\b{u}^*(\b{x}_0)+O\left(\frac{|\b{u}^*(\b{x}_0)|}{h^2}\right),\label{util}\\
\bm{\tilde{\Omega}}=\frac{1}{2}\nabla \times \b{u}^*(\b{x}_0)+\Gamma \b{\hat{e}}\times \b{E}^*(\b{x}_0)\cdot \b{\hat{e}}+O\left(\frac{|\b{u}^*(\b{x}_0)|}{h^3}\right),
\end{gather}
where $\Gamma=(1-\gamma^2)/(1+\gamma^2)$, $\gamma$ is the body aspect ratio, and $\b{E}^*=(\nabla \b{u}^*+\nabla (\b{u}^*)^T)/2$ is the symmetric rate of strain tensor. The full expressions for $\tilde{\b{u}}$ and $\bm{\tilde{\Omega}}$ are included in Appendix B, and we will use these full expressions in numerical simulations, but for the sake of mathematical tractability we now also consider the leading order dynamics assuming $h/A \ll 1$. Caution must be taken here, as we are expanding expressions valid for $1/h^2 \ll 1$ (see Eq.~\ref{util}) in the small parameter $h/A$. In other words, it is important that $A \gg 1$ for what follows (the colloid must be much larger than the swimmer). 

Inserting the expressions for $\b{\tilde{u}}$ and $\bm{\tilde{\Omega}}$ into Eqs.~\eqref{hdot_general}-\eqref{tdot_general}, we find the following model equations for the dynamics,
\begin{gather}
\frac{dh}{dt}=\sin(\theta)-\frac{3 \alpha}{8 h^2}(1-3 \sin^2 \theta),\label{hdot0}\\
\frac{d\theta}{dt}=\frac{1}{A} \cos \theta-\frac{3\alpha }{64 h^3} \left[4-\Gamma (3-\cos 2 \theta )\right] \sin 2 \theta.\label{tdot0}
\end{gather}
Eqs.~\eqref{hdot0}-\eqref{tdot0} in the limit as $A\rightarrow \infty$ have been used by other authors to study self-propulsion near infinite plane walls \cite{btbl08,sl12,ddcgg11}. We observe that the leading order variation in the dynamics from the infinite-wall case is due solely to the geometric effect, and not to variations in the hydrodynamic effects.  Note that the far-field hydrodynamic approximations of swimming bodies were found to give surprisingly accurate results for motion near an infinite plane wall, as compared to solutions of the full Stokes equations for Janus swimmers of varying eccentricity, for motion as close as fractions of a body length away from the surface \cite{sl12}. 

\section{ Hydrodynamic collision: entrapment and scattering}\label{sec: entrapment and scattering}

Previous studies of self-propulsion near infinite plane wall surfaces have shown that  pushers  $(\alpha>0)$ swimming  nearly parallel to the wall are attracted to a planar surface by a passive hydrodynamic interaction. Pullers ($\alpha<0$), meanwhile, are repelled in this configuration. With these effects in mind, we now look to the case of a finite colloid size. Note that in this deterministic setting, the \tcb{swimmer is confined to the plane spanned by the swimming director and the line of centers between the swimmer and the colloid; coordinates can be defined so that the swimmer is confined to the $x-y$ plane, for instance.}

\begin{figure}
\begin{center}
\includegraphics[width=.4\textwidth]{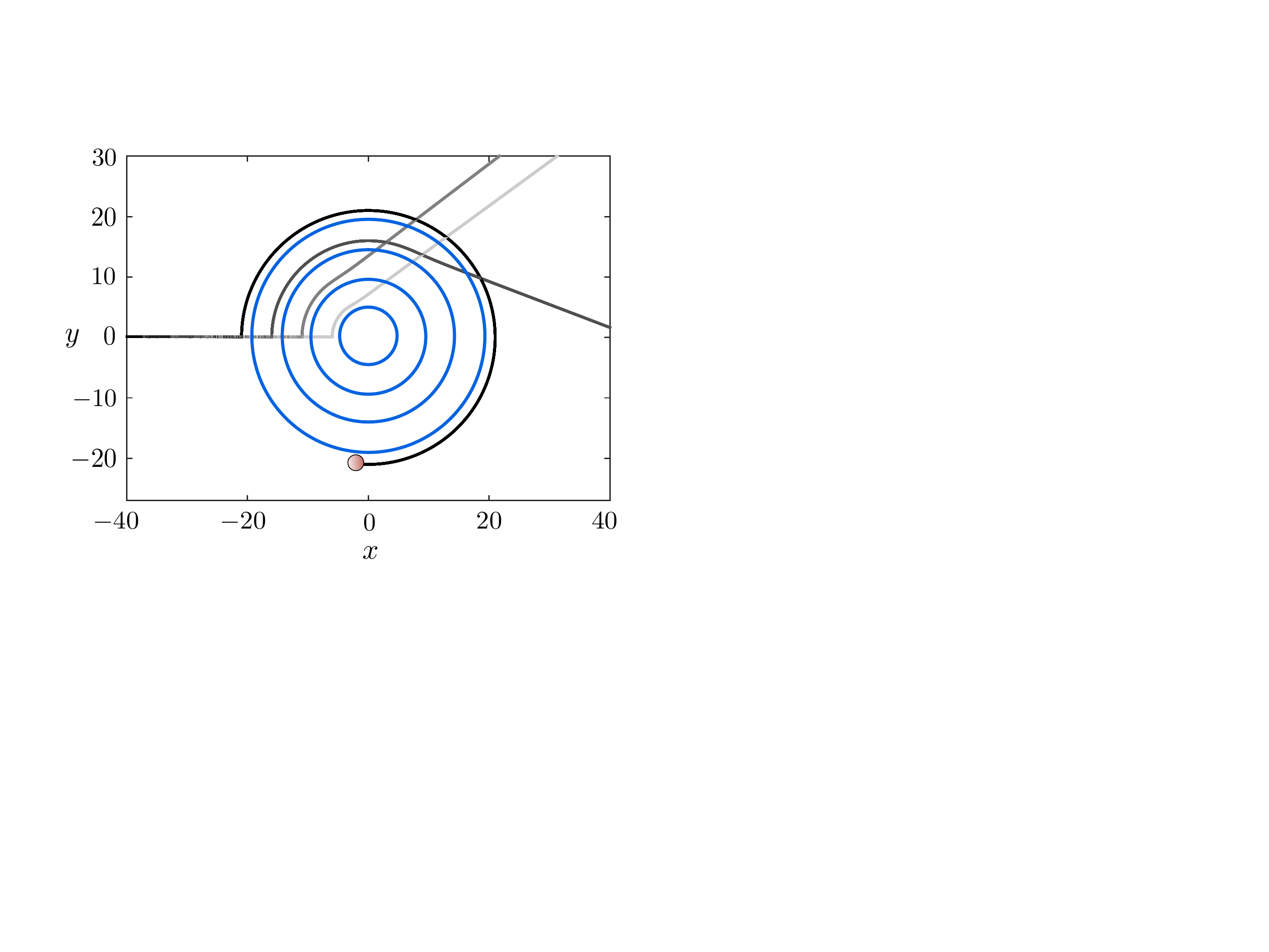}
\caption{A spherical dipole pusher $(\alpha=0.8)$ with initial position $\b{x}_0=-40\b{\hat{x}}+0.1\b{\hat{y}}$ and orientation $\b{\hat{e}}(0)=\b{\hat{x}}$ swims towards colloids of radius $A=5$, $10$, $15$, and $20$ in the time $t:0\to 120$. The critical colloid size for entrapping a swimmer with $\alpha=0.8$ is $A_{\rm c}\approx 15.1$. The simulations are produced by integrating numerically  the full system of Eqs.~\eqref{Mainut}-\eqref{MainOt}.}
\label{Bullseye}
\end{center}
\end{figure}

We begin by investigating numerically the dynamics of a dipole swimmer using the complete far-field approximation (Eq.~\eqref{hdot_general} with no assumption that $h/A \ll 1$, as described in Appendix B). We show in Fig.~\ref{Bullseye}   the trajectories of a spherical pusher with strength $\alpha=0.8$ and initial position $\b{x}_0=-40\b{\hat{x}}+0.1\b{\hat{y}}$ and orientation $\b{\hat{e}}(0)=\b{\hat{x}}$ as it swims towards colloids centered about the origin of varying sizes. For small colloid sizes, $A=5$ and $A=10$, the swimmer makes hard contact with the sphere, then turns and travels along the colloid until escaping  
from the surface. The colloid of size $A=15$ makes escape more difficult but the swimmer is eventually able to propel freely away from the sphere. However, for all colloid sizes larger than $A\approx 15.1$, the colloid captures the swimmer. The swimmer is trapped in a periodic orbit and endlessly propels past the surface of the colloid, as shown for the case $A=20$.

More generally, the critical colloid size for entrapment, denoted by $A_{\rm c}$, depends on the dipole strength $\alpha$ and the aspect ratio of the swimmer. The critical colloid size for entrapping a spherical pusher or puller is shown in Fig.~\ref{Ac_compare} along with predictions to be described in the following section. The size of the colloid is found to scale as $1/\alpha^2$ when $\alpha>0$ (for pushers) and as $1/|\alpha|$ for pushers.

\subsection{Estimating the critical trapping radius}

\begin{figure}
\begin{center}
\includegraphics[width=.44\textwidth]{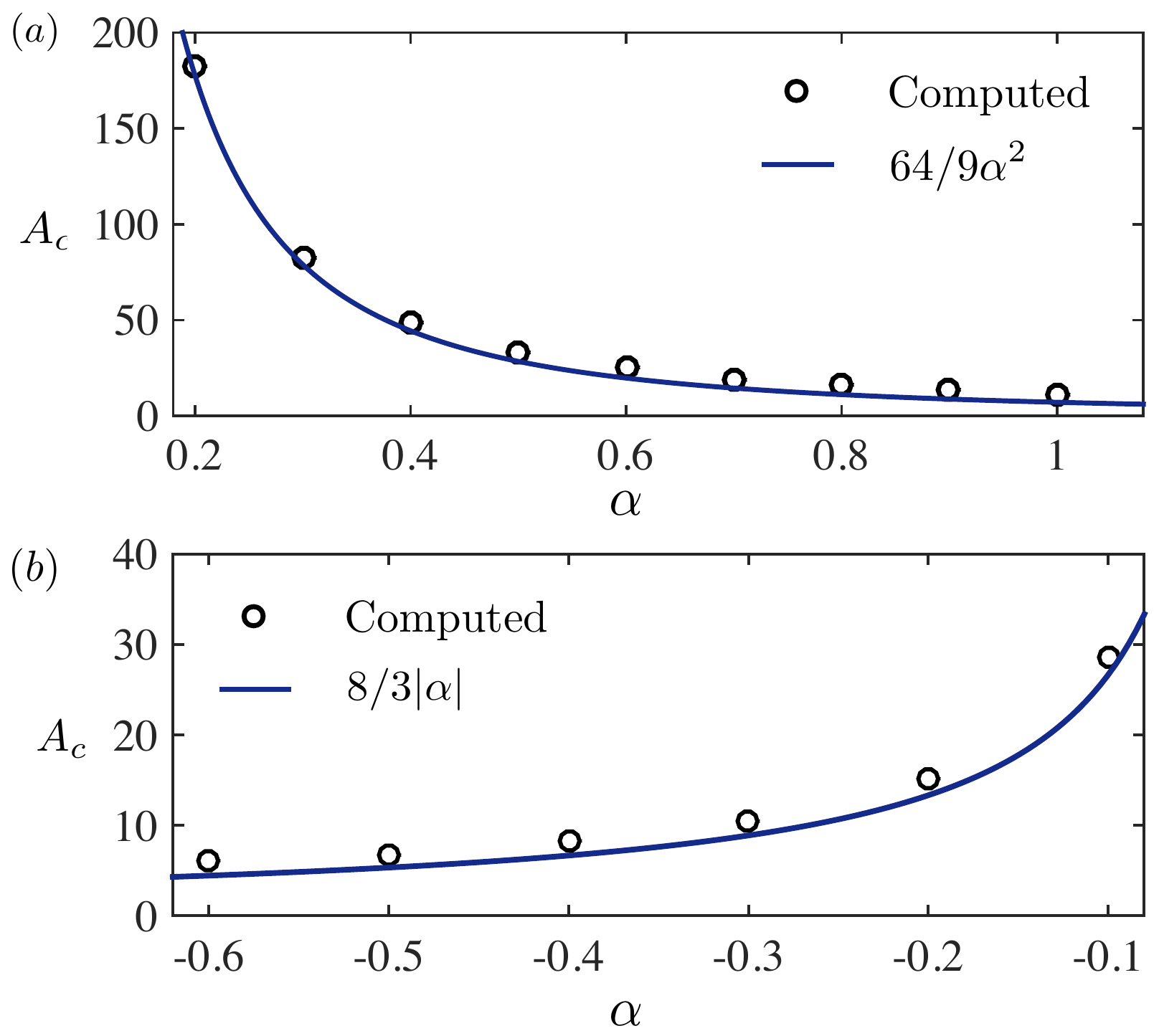}
\caption{(a) The critical colloid size for entrapment, $A_{\rm c}$, as a function of the dipole strength $\alpha$ for a spherical ``pusher'' swimmer. Values computed using initial position $\b{x}_0=(A+1)\b{\hat{y}}$ and $\Theta=0$ are shown as symbols, and the prediction $A_{\rm c}=64/9\alpha^2$ as a solid line. The theory is strongest for smaller dipole strengths and larger colloid sizes, where the escape angle is smaller and the linearized equations are more accurate. (b) The same, for puller swimmers, along with the theoretical prediction of $A_{\rm c}=8/3|\alpha|$.}
\label{Ac_compare}
\end{center}
\end{figure}
One of the primary goals of this paper is to estimate the relationship between the dipole strength, $\alpha$, and the critical colloid size 
$A_{\rm c}$.  Linearizing Eqs.~\eqref{hdot0}-\eqref{tdot0} about $\theta=0$ (swimming parallel to the colloidal surface), pushers are found to be attracted to the surface and pullers are repelled from the surface, just as in the infinite wall case \cite{btbl08}. However, unlike the dynamics near a plane wall, for finite colloid size $A$ we now have $\dot{\theta}>0$ when $\theta=0$ as a consequence of the topographical curvature. Hence, $\theta=0$ is no longer an equilibrium pitching angle and the body cannot swim parallel to the surface for any sustained period of time. Linearizing the system about $\theta=0$, 
\begin{gather}
\frac{dh}{dt}= \theta-\frac{3 \alpha}{8 h^2}+O\left(\theta^2\right),\label{hdotapprox0}\\
\frac{d\theta}{dt}=\frac{1}{A}-\frac{3\alpha(2-\Gamma)}{16 h^3}\theta+O\left(\theta^2\right),\label{tdotapprox0}
\end{gather}
we find an equilibrium solution $h^\star=\left(9\alpha^2 A(2-\Gamma)/4\right)^{1/5}/2$ and $\theta^*=\left(3\alpha/[4A^2(2-\Gamma)]\right)^{1/5}$. 

{Let us focus first on the pusher case}. Here we see that $\theta^*>0$ for $\alpha>0$. The normalized equilibrium distance $h^\star/A$ decreases with increasing $A$ as expected (a larger sphere gives a larger hydrodynamic attraction), but surprisingly increases with  $\alpha$ due to the effect of the dipole strength on the rotation rate. However, it is not difficult to show that this solution is not asymptotically stable, and instead corresponds to a saddle point in the dynamics. Instead, given the nature of the hydrodynamic attraction, we expect  hydrodynamic {capture to be achieved when there is a balance between  hydrodynamics and some other physical repulsion, which we model here as an effective hard-core interaction. We can then} estimate a criterion for entrapment by fixing $h=\bar{h}$ when the swimming body is in contact with the colloid ($\bar{h}=1$ for a spherical swimmer). We recall that when hard contact is established, we still allow the pitching angle $\theta$ to evolve. Consistent with the linearization about small $\theta$ we set $h=\bar{h}\approx \gamma$ in Eq.~\eqref{tdotapprox0}, and we infer the pitching angle for which the geometric and hydrodynamic effects are in balance:
\begin{gather}
\theta^\star=\frac{16 \bar{h}^3}{3 A \alpha(2-\Gamma)}\cdot \label{tst}
\end{gather}
We note that $\theta^\star$ vanishes in the infinite-wall or infinite dipole strength limit, $A\alpha \rightarrow \infty$. Recalling that $\Gamma=(1-\gamma^2)/(1+\gamma^2)$, the predicted equilibrium angle is monotonically increasing in the swimmer aspect ratio $\gamma$ from a value of zero for a very slender swimmer ($\gamma=0, \Gamma=1$) to a positive value of $8/(3 A\alpha)$ for a spherical pusher $(\gamma=1,\Gamma=0)$. Physically, a slender swimmer is able to draw nearer to the colloid, where the hydrodynamic attraction is more significant, thereby making the surface of the colloid hydrodynamically more akin to an infinite plane wall.

\begin{figure}
\begin{center}
\includegraphics[width=.45\textwidth]{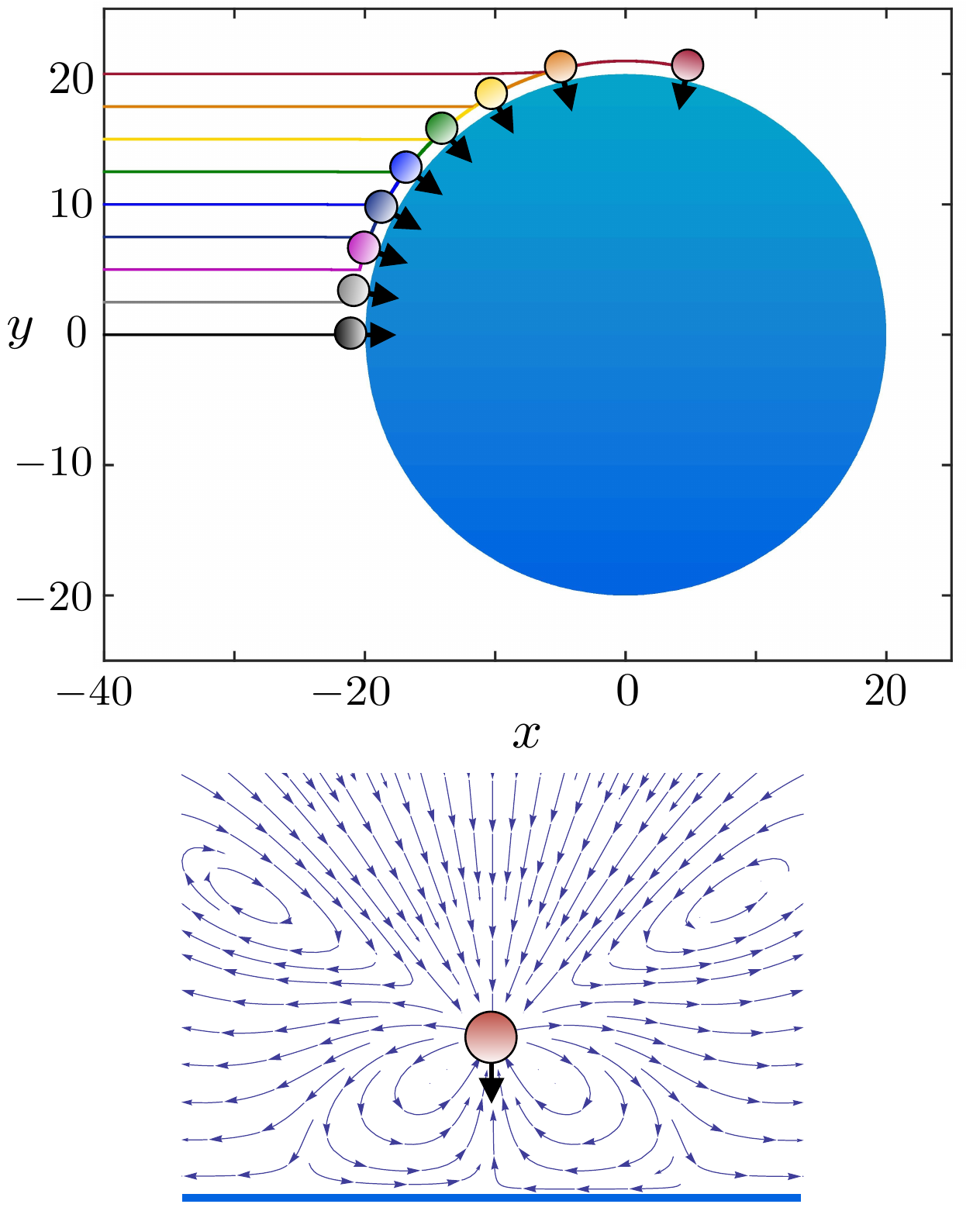}
\caption{(Top) Pullers ($\alpha=-0.8$) swim towards a sphere of size $A=20$, released from initial points $\b{x}_0=-40\b{\hat{x}}+2.5 j\b{\hat{z}}$ where $j$ ranges from 0 to 8, and angle $\Theta_0=0$ in the lab frame. The trajectories are computed for $t:0\to 100$, and in each case the swimmer comes to a steady equilibrium at the location shown, generally much earlier than $t=100$. \tcb{(Bottom) The flow-field directions (flow strength is not indicated) created by a puller swimming towards the surface of a large colloid.}}
\label{Pullers}
\end{center}
\end{figure}

\begin{figure*}[htbp]
\begin{center}
\includegraphics[width=.96\textwidth]{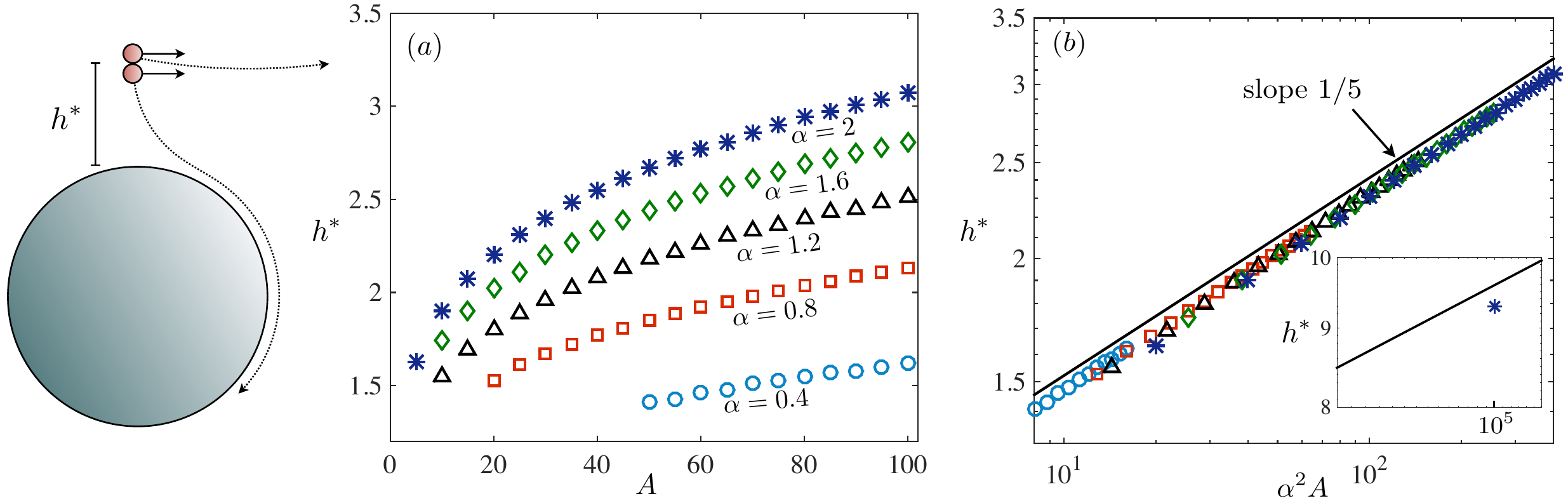}
\caption{Basin of attraction. For a spherical swimmer placed initially parallel to the surface, $\theta(0)=0$, $h^\star$ denotes the critical initial distance from the colloid above which the particle escapes, and below which entrapment ensues. (a) The critical initial distance for a selection of dipole strengths, shown where the colloid size is larger than the critical size for entrapment. (b) The curve collapses upon plotting against $\alpha^2 A$ to a power law scaling with exponent $1/5$. The solid line is the prediction from Eq.~\eqref{hstar}. \tcb{(Inset) The trend continues over five orders of magnitude in $\alpha^2A$.}}
\label{Basin}
\end{center}
\end{figure*}

This equilibrium pitching angle may now be used to propose a criterion for hydrodynamic capture. The question of escape now reduces to determining whether or not $\dot{h}$ in Eq.~\eqref{hdot0} is positive when $\theta=\theta^*$. A positive value of $\dot h$ indicates that the swimmer moves away from the surface. Using the same linearization about $\theta=0$ and inserting $\theta^*$ above into Eq.~\eqref{hdotapprox0} (with $h=\bar{h}$ fixed) we obtain a critical colloid size $A_{\rm c}$ for which $\dot{h}=0$:
\begin{gather}\label{Acrit} 
A_{\rm c}=\frac{128 \bar{h}^5}{9\alpha^2 (2-\Gamma)}\cdot
\end{gather}
For colloid sizes $A>A_{\rm c}$ we predict hydrodynamic capture; conversely for $A<A_{\rm c}$ the hydrodynamic attraction cannot trap the swimmer, which will continue to rotate until it reaches a critical pitching angle $\theta_{\rm e}$ for escape (the angle for which $\dot{h}$ becomes positive),
\begin{gather}
\theta_{\rm e}=\frac{3\alpha}{8\bar{h}^2},
\end{gather}
which is notably independent of the colloid size $A$. For a spherical swimmer we therefore predict a critical colloid size for capture of 
\begin{gather}
A_{\rm c}=\frac{64}{9\alpha^2}.
\end{gather}

Is this capture criterion borne out by full numerical integration of Eq.~\eqref{hdot_general}? Returning to Fig.~\ref{Ac_compare}a we find a very close agreement between this criterion and the numerically determined critical colloid sizes for a range of dipole strengths with the estimate above. The theory is strongest for smaller dipole strengths and larger colloid sizes, where the escape angle is smaller and the linearized equations are more accurate.

Pullers, however, act very differently near the colloid. For a spherical puller ($\alpha<0, \Gamma=0$), upon examination of Eq.~\eqref{tdot0} we see that the angle for which the swimmer is directly facing the surface and is motionless there, $\theta=-\pi/2$, is linearly stable as long as the colloid is of size $A=8/(3|\alpha|)$ or larger, which is considerably smaller than the colloid size required to trap a pusher for the range of $\alpha$ most relevant to microorganisms. Figure~\ref{Pullers} shows the trajectories of non-interacting pullers with $\alpha=-0.8$ swimming towards a sphere of size $A=20$. In each case, the swimmer quickly reaches a steady equilibrium at the location shown in Fig.~\ref{Pullers}.  We should therefore expect to see dramatic entrapment of such swimmers on trajectories which bring the swimmer almost directly into contact with the colloid. The ``suction'' in the direction of locomotion requires such a direct impact; an oblique interaction would result in a hydrodynamic repulsion, as depicted by the flow field shown in Fig.~\ref{FlowFields} but with the sign of the velocity everywhere reversed. The estimate of the critical colloid size is compared again to the results of the numerical simulations in Fig.~\ref{Ac_compare}b, and once again we obtain excellent agreement.

\subsection{Basin of attraction}

We next investigate the  basin of attraction, i.e.~the domain in space over which the particle is eventually captured by the colloid. In the regime studied, with $h/A \ll 1$ and $\alpha=O(1)$, the basin of attraction has a radius not much larger than the colloid itself. For instance, even with $A=200$ and $\alpha=0.8$, if a spherical swimmer is initially placed parallel to the surface, the initial distance from the colloid below which the body is trapped is approximately $h=2.5$, smaller than three body lengths away. For $A=20$, the value is smaller still and the spherical swimmer in this case must be placed closer than $h=1.5$ from the surface, only a percentage of its size away from the colloid. In general, we therefore expect that hydrodynamic trapping may be a strong effect, but only for particles that are on a trajectory that leads to a direct contact with the obstacle. 

In Fig.~\ref{Basin}a we show the initial value of $h$, with $\theta(0)=0$ (initial swimming is parallel to the surface) such that the swimmer is captured at the colloid surface. This basin depth, defined by $h=h^\star$, naturally increases  with both increasing dipole strength $\alpha$ and colloid size $A$. Here again, as in the estimation of the critical colloid sizes leading to {capture}, the quantity $\alpha^2 A$ is found to play a critical role. Plotting $h^\star$ as a function of $\alpha^2 A$ reveals a collapse of the data to a single mastercurve, $h^\star\approx h^\star(\alpha^2 A)$, for almost the complete range of $A$ and $\alpha$ considered, as shown in Fig.~\ref{Basin}b.

\begin{figure*}
\begin{center}
\includegraphics[width=\textwidth]{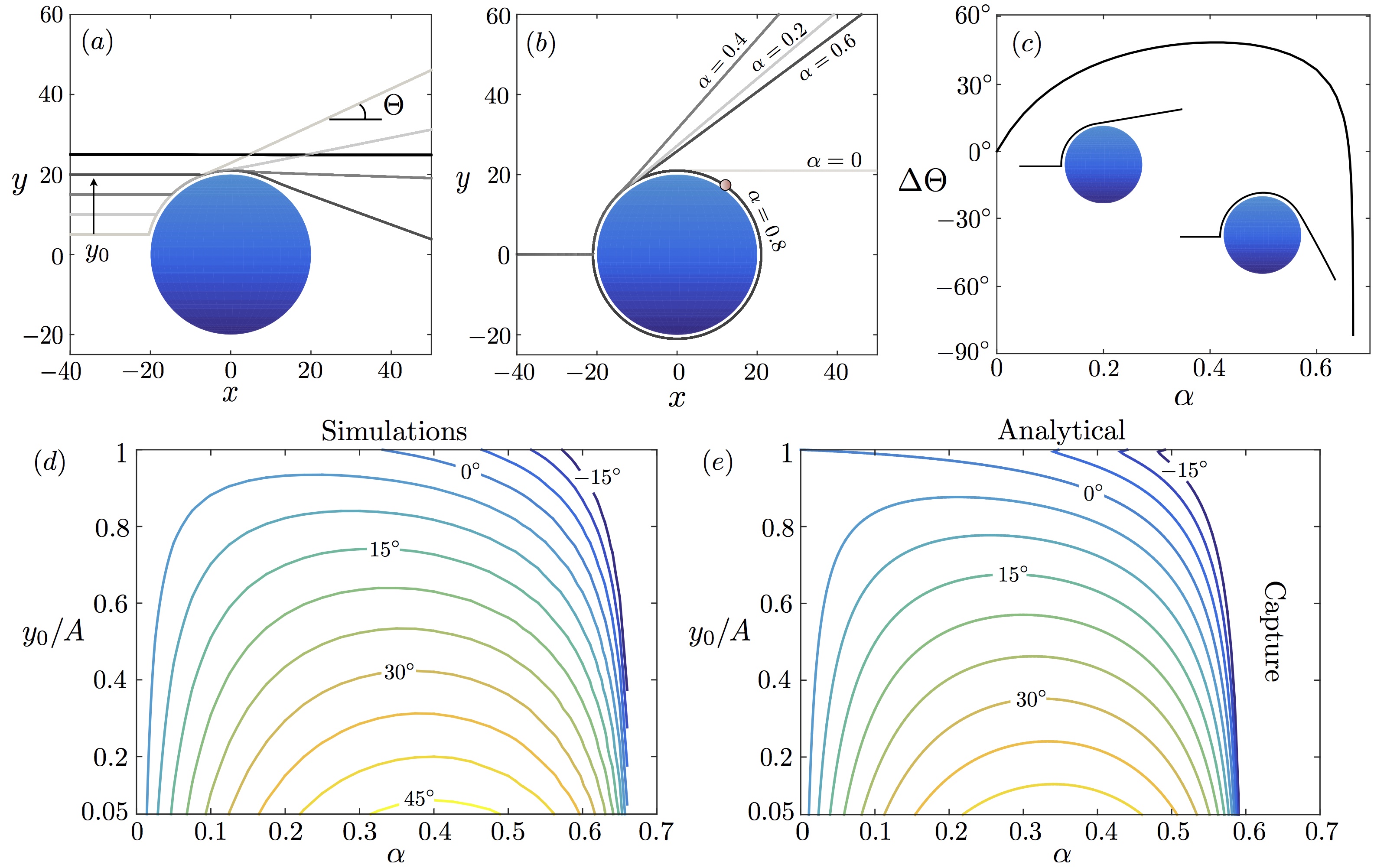}
\caption{Scattering of a spherical swimmer with initial position $\b{x}_0=-40\b{\hat{x}}+y_0\b{\hat{y}}$ and orientation $\Theta_0=0$ by a spherical colloid of fixed size $A=20$. (a) Fixing the dipole strength to $\alpha=0.6$, the scattering angle $\Delta\Theta$ is non-monotonic in the impact parameter $y_0$. (b-c) Fixing the impact parameter to $y_0=0.1$, the scattering angle is also non-monotonic in the dipole strength $\alpha$. Swimmers with $\alpha>0.67$ become hydrodynamically bound to the colloid, corresponding to a singularity in the scattering angle. For $\alpha$ extremely close to its critical value the swimmer may wind around the colloid multiple times before departing from the surface (see Fig.~\ref{WindingNumber}). (d) The scattering angle for a range of impact parameters and dipole strengths, from simulations. Contours are shown for multiples of $5^\circ$. (e) The analytical prediction from Eq.~\eqref{fullscattering}.}
\label{TotalScattering}
\end{center}
\end{figure*}

In order to estimate theoretically  the basin depth, $h^\star$, we consider a spherical swimmer, $\Gamma=0$, and perform a Taylor expansion of the dynamics at small times, $h(t)=h_0+h_1 t+h_2 t^2+\dots$, $\theta(t)=\theta_1 t+\theta_2 t^2+\dots$. Inserting these expansions into Eqs.~\eqref{hdotapprox0}-\eqref{tdotapprox0} and matching terms of like powers of $t$, we find
\begin{gather}
h(t)=h_0-\frac{3\alpha}{8 h_0^2}t+\left(\frac{1}{2A}-\frac{9\alpha^2}{64 h_0^5}\right)t^2+\dots,\\
\theta(t)=\frac{t}{A}-\frac{3\alpha}{16 A h_0^3} t^2+\dots.
\end{gather}

Using the expression for $h(t)$ up to quadratic terms in $t$, the distance from the colloid is seen to be minimal when $t_{\rm min}=12 A \alpha h_0^3/(32 h_0^5-9\alpha^2 A)$. Setting this value to unity would seem to distinguish whether the swimmer makes eventual contact with the colloid, but this results in a poor approximation. Instead, we look to the equation for $\theta(t)$ at this moment in time. The angle $\theta(t_{\rm min})=3\alpha/8 h(t_{\rm min})^2$ is an unstable fixed point for the dynamics as  noted earlier (see Eq.~\ref{tdotapprox0}). For a value $\theta(t_{\rm min})$ smaller than this critical value the swimmer will collapse towards the colloid, while for larger values the swimmer will escape. Using the quadratic expressions in time above, and setting $\theta(t_{\rm min})=3\alpha/8 h(t_{\rm min})^2$ as the boundary case, we arrive at an equation for the initial height $h_0$, {which approximates  the critical capture distance $h^\star$,
\begin{gather}
h^\star=\rho^{1/5}(\alpha^2 A)^{1/5},\label{hstar}
\end{gather}
where the prefactor $\rho^{1/5}\approx0.96$ corresponds to the only real zero of a third order polynomial, $16384 \rho^3-24192 \rho^2+10611 \rho-1458=0$. This analytical prediction is in excellent agreement  with the results from the numerical simulations (solid curve in Fig.~\ref{Basin}b). We stress that the scaling $(\alpha^2 A)^{1/5}$, which reflects the subtle interplay between self-propulsion, contact, and hydrodynamic reorientation, could not have been anticipated from a dimensional analysis alone. 

\subsection{Scattering by a spherical obstacle}

Now that we have gained intuition about the physical mechanisms responsible for  swimmer capture, we lay out a comprehensive description of the scattering process in the case of a spherical pusher swimming toward a spherical obstacle. Figure~\ref{TotalScattering} provides a general picture of the scattering dynamics, where we fix the colloid size to $A=20$. The initial orientation angle in the lab frame is $\Theta_0=\sin^{-1}(\b{\hat{x}}\cdot \b{\hat{e}}(0))=0$, and the swimmer is initially located at a position $\b{x}_0=-40\b{\hat{x}}+y_0 \b{\hat{y}}$, where $y_0$ is called the impact parameter.

Figure~\ref{TotalScattering}a shows the trajectories of a swimmer with $\alpha=0.6$ near a colloid of size $A=20$, where we vary the impact parameter $y_0$. The interaction of the swimming body with the spherical surface need not be long lived in order for the swimmer to be redirected dramatically.  The amount of time spent in close contact with the sphere decreases monotonically with increasing $y_0$. In contrast, the scattering angle displays non-monotonic variations with the impact parameter, as seen in Fig.~\ref{TotalScattering}b. Of particular note, the impact with $y_0=A$ has only a brief period of contact with the sphere, but the hydrodynamic attraction to the surface is sufficiently strong to induce a strong scattering of the swimming trajectory, which results in a scattering angle as large as $\Delta \Theta\approx-18^\circ$. The swimmer for which $y_0=A/4$, on the other hand, interacts with the colloid for a longer period of time, but it departs from the surface in such a way as to result in a positive change in the swimming angle, even though the interaction is much more dynamic. Comparing all four cases shown it is clear that the scattering angle can be positive or negative, small or large, and is rather sensitive to the swimmer's trajectory of approach. 

Furthermore, we observe that the scattering angle is also non-monotonic in the dipole strength. In Fig.~\ref{TotalScattering}c we plot the trajectories of spherical pushers of varying dipole strength $\alpha$ through their interactions with a colloid of radius $A=20$. The case $\alpha=0$ (no hydrodynamic interactions) results in no change in the swimming director, only a lateral translation in space as the swimmer slowly pushes past the spherical obstacle. The final swimming direction is not a simple monotonic function of $\alpha$, as shown in Fig.~\ref{TotalScattering}d, and a singularity appears in the scattering angle as $\alpha$ approaches a critical value for entrapment.

The variation of the deflection angle as a function of the impact parameter $y_0$ is shown in Fig.~\ref{TotalScattering}e for the same dipole strengths as in Fig.~\ref{TotalScattering}c (in which the impact parameter is fixed to $y_0=0.1$). A rapid transition is observed for impact parameters very near to $A$. The scattering angle is nearly zero for values $h_0/A$ not much larger than one (recall the small depth of the basin of attraction), indicating that the effective cross-section of the colloid is not significantly different from its diameter even though hydrodynamic interactions are long-ranged. The capture of the swimmer is again clearly revealed by the singularity in the scattering plot for $\alpha=0.8$.

\subsection{Estimating the scattering angle}
\begin{figure}
\begin{center}
\includegraphics[width=.45\textwidth]{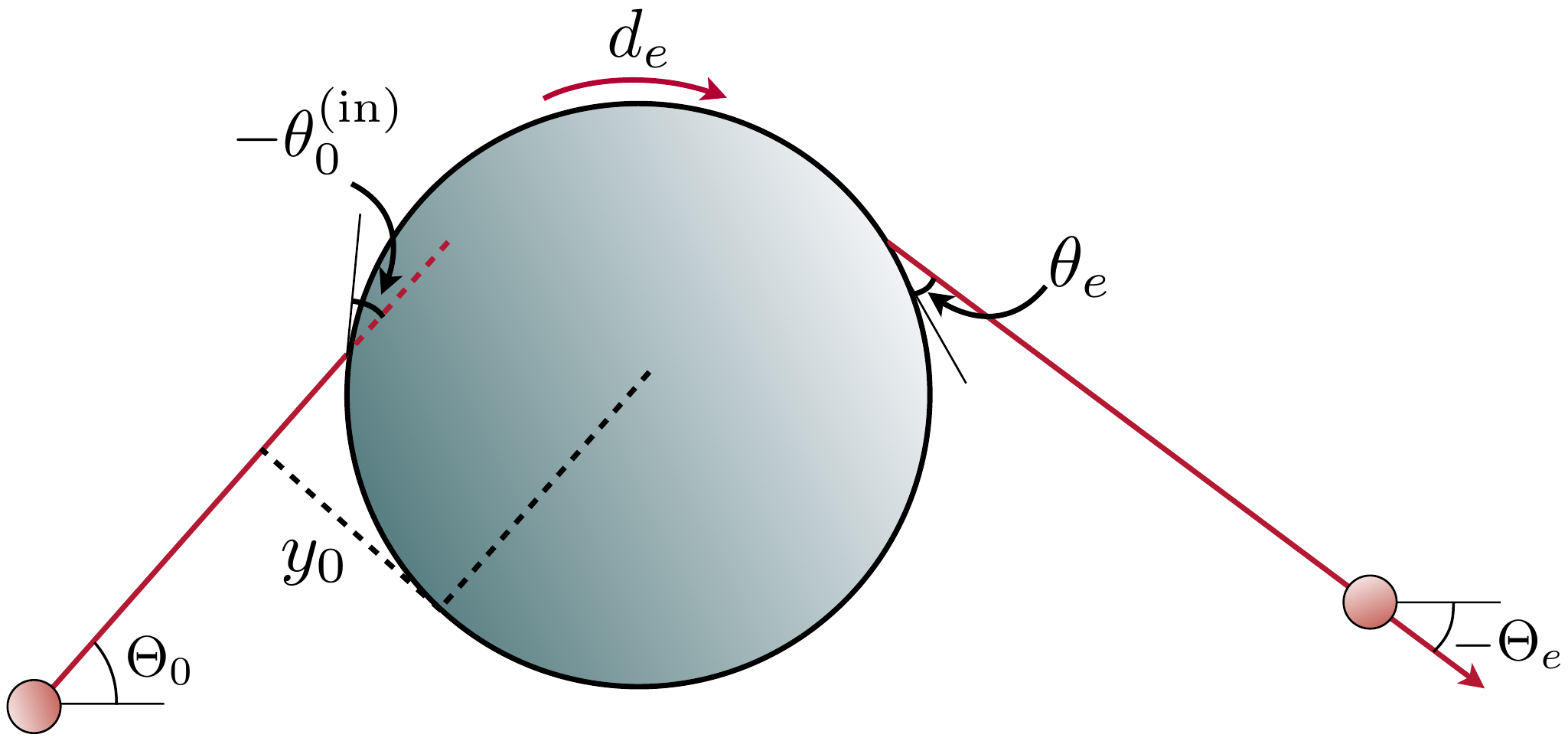}
\caption{Scattering interaction of a swimming body with a colloid of sub-critical size for entrapment, $A<A_{\rm c}$. The impact angle with no hydrodynamic interaction is denoted by $\theta^{\rm{(in)}}_0$, the impact parameter is $y_0$, the distance travelled along the surface by $d_e$, and the escape angle by $\theta_{\rm e}$. The total scattering angle in the lab frame is given by $\Delta \Theta = \Theta_e-\Theta_0$.}
\label{ScatteringIllustrated}
\end{center}
\end{figure}

We now proceed to estimate the scattering angle of a spherical pusher that impacts a colloid of sub-critical size for entrapment, $A<A_{\rm c}$.  In order to do so we decompose the scattering process into three steps (see Fig.~\ref{ScatteringIllustrated}): (i) the approach toward the colloid during which hydrodynamic interactions modify the orientation of the swimmer at a distance, (ii) the sliding of the swimmer over the colloid surface, and (iii) the escape during which the hydrodynamic interactions act again at a distance. 

The approach (step i) may be described using Eqs.~\eqref{hdotapprox0}-\eqref{tdotapprox0}. We define the contact time as $t=0$, at which point the body is oriented at an angle $\theta_0$, assumed to be small, and $h=1$. Before impact, approximating the distance from the surface as $h=1+\theta_0 t$ for $t<0$ and that $\theta\approx \theta_0$, then the body rotation may be estimated by integrating the hydrodynamic effect on rotation alone (ignoring the geometric part of Eq.~\eqref{tdotapprox0}),
\begin{gather}
\Delta\Theta_{-\infty\rightarrow 0}= -\frac{3\alpha}{8}\int_{-\infty}^0 \frac{\theta_0}{[1+\theta_0 t]^3}\,dt=\frac{3\alpha}{16}\label{Delta_Theta_Minus_Inf}.
\end{gather}
Therefore, with the unimpeded impact angle illustrated in Fig.~\ref{ScatteringIllustrated} given by $\theta_0^{\rm{(in)}}=\sin^{-1}(y_0/A)-\pi/2$, then the adjusted impact angle is estimated as $\theta_0=\theta_0^{\rm{(in)}}+3\alpha/16$. 

Next we describe the sliding motion of the swimmer in contact with the colloid (step ii). Integrating Eq.~\eqref{tdotapprox0} with initial condition $\theta(0)=\theta_0$, we find
\begin{gather}
\theta(t)=\theta^*+\left(\theta_0-\theta^*\right)e^{-3\alpha t/8},\label{theta_scattering}
\end{gather}
where $\theta^*=8/(3\alpha A)$ is the fixed point of $\dot{\theta}$ when $h=1$. The time at which $\theta$ reaches the escape angle $\theta_{\rm e}=3\alpha/8$ is therefore 
\begin{gather}\label{te}
t_e=\frac{8}{3\alpha} \log\left(\frac{1-\theta_0/\theta^*}{1-A/A_{\rm c}}\right),
\end{gather}
with $A_{\rm c}=64/9\alpha^2>A$, and the distance traveled is approximated simply by $d_e=t_e$. When the swimmer is in contact with the colloid, the dynamics of $\Theta(t)$ is given generally by $\Theta_t = \theta_t-\cos \theta(t) /A\approx  \theta_t-1/A$. Integrating from $t=0$ to $t=t_e$, the variation in the swimmer's orientation angle while the swimmer slides along the surface is 
\begin{gather}
\Delta \Theta_{0\rightarrow e}=(\theta_{\rm e}-\theta_0)-\frac{d_e}{A}\cdot \label{Delta_Theta_0}
\end{gather}

Finally, as the swimmer escapes from the colloid surface (step iii) we have the initial conditions $h(t_e)=1$ and $\theta(t_e)=\theta_{\rm e}$ which set the initial conditions of Eqs.~\eqref{hdotapprox0}-\eqref{tdotapprox0}. Once again carrying out a Taylor expansion for small time, we find for $t>t_e$ that $h(t)=\tilde{h}(t-t_e)+O((t-t_e)^5)$, where
\begin{gather}
\tilde{h}(t)=1+\left(\frac{1}{A}-\frac{1}{A_{\rm c}}\right)\left(\frac{1}{2}t^2+\frac{\alpha }{16}t^3+\frac{9\alpha^2}{256}t^4\right).
\end{gather}
Again assuming that $\dot{\theta}$ is small so that here $\theta \approx \theta_{\rm e}=3\alpha/8$, then the remainder of the body rotation is also found by integrating numerically only the hydrodynamic effect on rotation,
\begin{gather}
\Delta\Theta_{e\rightarrow \infty}= -\frac{9\alpha^2}{64}\int_0^\infty \frac{dt}{\tilde{h}(t)^3}\label{Delta_Theta_Inf}.
\end{gather}
That this expression is negative indicates that the hydrodynamic interaction causes the swimmer to rotate back towards alignment with the colloid surface after departure.

Combining steps (i-iii), we obtain the total scattering angle
\begin{gather}
\Delta \Theta = \frac{3\alpha}{16}+\frac{\pi}{2}-\sin^{-1}\left(\frac{y_0}{A}\right)-\frac{t_e}{A} -\frac{9\alpha^2}{64}\int_0^\infty \frac{dt}{\tilde{h}(t)^3},\label{fullscattering}
\end{gather}
with $t_e$ given in Eq.~\eqref{te} using $\theta_0=\sin^{-1}(y_0/A)-\pi/2+3\alpha/16$. In the limit of no hydrodynamic interaction with the colloid, $\alpha\to 0$, the expression above returns zero, as expected. Fixing the colloid size to be $A=20$, the scattering angle as a function of the impact parameter $y_0$ and the dipole strength $\alpha$ from the estimate above is shown in Fig.~\ref{TotalScattering}e, alongside the values determined by numerical simulations in Fig.~\ref{TotalScattering}d. We observe a close agreement between the two, with the prediction systematically overestimating the scattering angle in this case by a few degrees.

\begin{figure}[t]
\begin{center}
\includegraphics[width=.45\textwidth]{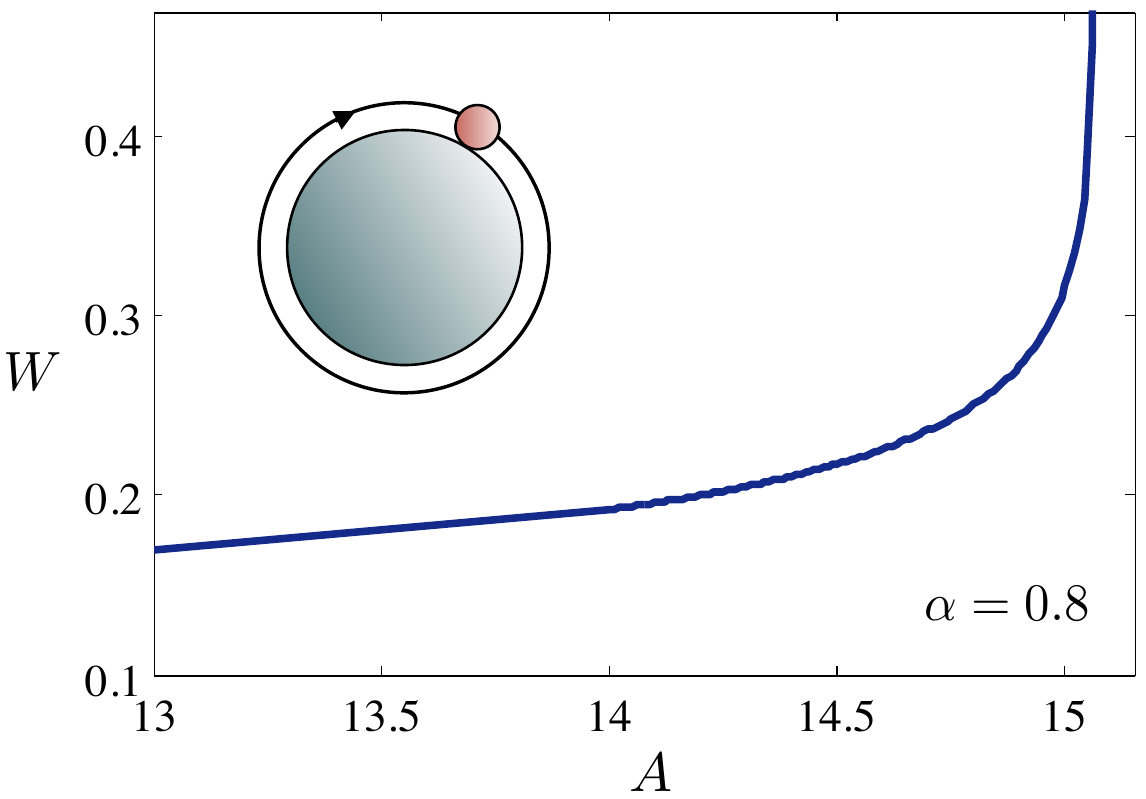}
\caption{\tcb{Fraction of an orbit traversed} around the spherical surface before escape from a colloid of subcritical size, $A<A_{\rm c}$.}
\label{WindingNumber}
\end{center}
\end{figure}

\begin{figure*}[t]
\begin{center}
\includegraphics[width=.8\textwidth]{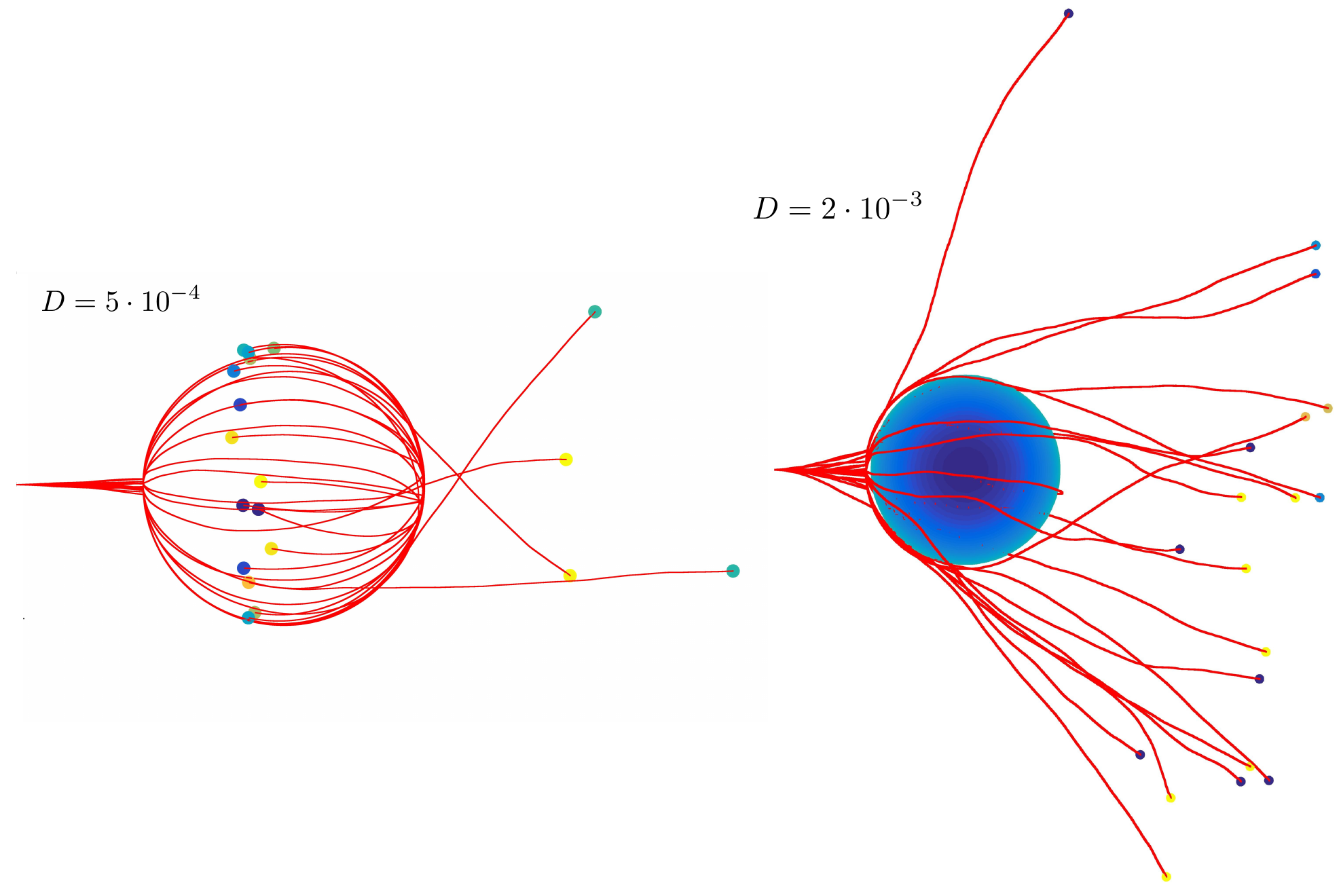}
\caption{Twenty instances of swimming trajectories for a pusher ($\alpha=0.8$) near a sphere of radius $A=20$ with initial position $\b{x}_0=-40\b{\hat{x}}$ and orientation $\b{\hat{e}}(0)=\b{\hat{x}}$, computed for $t:0\to 120$. (Left) The dimensionless diffusion constant is $D=5\cdot 10^{-4}$ and many of the instances remain hydrodynamically bound at $t=120$. Color coding / shading indicates the final position along the $z$-axis with darker swimmers coming out of the page and lighter swimmers going into the page. The colloid boundary may be inferred. (b) The same, but with a larger diffusion constant, $D=2\times 10^{-3}$. In this case none of the swimmers are bound to the colloid at $t=120$.}
\label{RandomSwimming}
\end{center}
\end{figure*}

An alternative way to quantify the swimmer-colloid interaction is to measure the number of orbits \tcb{(or fraction of an orbit)} around the colloid travelled by the swimmer before escape, given by the ratio $W=d_e/(2\pi A)\approx t_e/(2\pi A)$. The result in Eq.~\eqref{te} suggests that the residence time is continuous in its rapid increase to infinity as $A\to A_{\rm c}$. However, due to the logarithmic dependence on $1-A/A_{\rm c}$, unless $A$ is extraordinarily close to $A_{\rm c}$ the swimmer will undergo only a partial orbit before departure. For a very rough bound, taking $\theta_0=-\pi/2$ and $\theta_{\rm e}=\pi/2$, and setting $A=A_{\rm c}(1-\ep)$ for some small positive $\ep$, then $W=\log(2/\ep-1)/[4(1-\ep)]$, so that even one full revolution around the colloid requires $\ep\leq 0.043$, or $A$ must be within $4\%$ of $A_{\rm c}$ for the swimmer to make one complete orbit around the colloid. Figure~\ref{WindingNumber} shows the \tcb{fraction of the orbit traversed}, computed for the simulations shown in Fig.~\ref{Bullseye}, which shows precisely this logarithmic singularity as $A$ approaches the critical colloid size, $A_{\rm c}$. 

\section{Fluctuation-induced escape from a colloidal trap}\label{sec: fluct}
The dynamics of swimming microorganisms are anything but smooth and deterministic. Whether because of thermal fluctuations (Brownian motion) or other complex biological behaviors (e.g., run-and-tumble locomotion of {\it E. coli}), randomness plays an important role in the trajectories of microorganisms and synthetic microswimmers. To evaluate the robustness of our findings for the deterministic problems studied in the previous section, we now consider the effects of fluctuations  on the interaction dynamics between the swimming body and the colloid. We confine our study to the case of pushers $(\alpha>0)$. 

To gain some intuition about the effects of random fluctuations, the full nonlinear model is solved with the addition of noise. We model the trajectory of a swimmer considering the effect of random forces and torques on the translational, and rotational dynamics by Langevin equations,
\begin{eqnarray}
\frac{d\b{x}}{dt}= \left(\b{\hat{e}}+ \tilde{\bf u}\right)+ \sqrt{6D}\,\bm{\eta}(t),\\ \label{Bswimmer_dimensionless}
\frac{d\hat{\b{e}}}{dt}= \left(\tilde{\bm{\Omega}}+\sqrt{4D_r} \, \bm{\eta}_{\rm R}(t)\right)\times \b{\hat{e}},\label{Bswimmer-angle} 
\end{eqnarray}
where $\b{\hat{e}}$ is the unit direction of swimming, and $\b{\tilde{u}}$ and $\bm{\tilde{\Omega}}$ are contributions from the hydrodynamic interaction with the colloid (\S\ref{sec: equations}). Forces and torques from thermal fluctuations are proportional to normalized Gaussian white noise in three-dimensions, $\bm{\eta}(t)$, and on a sphere, $\bm{\eta}_{\rm R}(t)$, where $\langle \eta_i(t) \eta_j(t')\rangle = \delta_{ij} \delta(t-t')$ and $\langle (\eta_{\rm R})_i(t) (\eta_{\rm R})_j(t')\rangle = \delta_{ij}\delta(t-t')$.

In an infinite viscous fluid, the dimensionless constants of translational diffusion, $D$ and rotational diffusion, $D_r$, are related by an application of the fluctuation-dissipation theorem and insertion of the mobilities of a sphere, so that $D_r = 3 D/4$, though this relation in general will depend on $h$. For this first exploration we will assume the relation $D_r=3D/4$.

\begin{figure*}[t]
\begin{center}
\includegraphics[width=\textwidth]{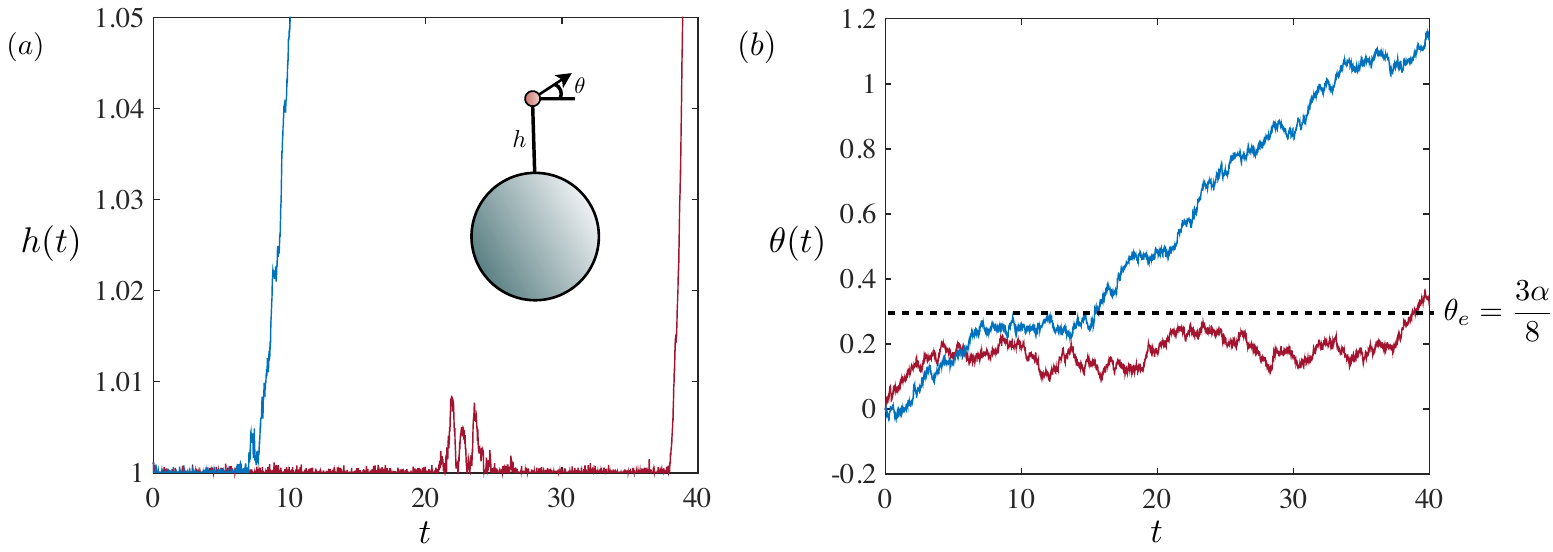}
\caption{(a) The distance to the colloid, $h(t)$, for two instances of swimmers with $h(0)=1.001$ and $\theta(0)=0$, and diffusion constant $D=2\times 10^{-3}$. The intermittency of near-surface swimming is due to translational Brownian fluctuations, and the hydrodynamic attraction rapidly returns the swimmer to the surface. (b) The local pitching angle, $\theta(t)$, for the same two instances as in (a). Eventual escape in this regime of $(A,\alpha,D)$ is due to $\theta$ nearing the deterministic escape angle, $\theta_{\rm e}=3\alpha/8$ for spherical swimmers, in concert with random fluctuations.}
\label{TrajectoriesA}
\end{center}
\end{figure*}

Using this framework we now show how noise allows microswimmers to escape hydrodynamic traps. We show in Fig.~\ref{RandomSwimming}  twenty instances of the swimming trajectory of a spherical swimmer with $\alpha=0.8$ near a colloid of size $A=20$, released from $\b{x}(0)=-40\b{\hat{x}}$ with initial orientation $\b{e}(0)=\b{\hat{x}}$. A forward Euler method is used to integrate the stochastic differential equations with time-step size $\Delta t= 0.001$. Simulating the dynamics in the time interval from 0 to 120, the first panel shows that in a few instances with $D=5\times 10^{-4}$ the swimmer makes contact with the colloid surface but then escapes, never to return, while many others remain trapped in this time interval. Meanwhile, the second panel shows the same swimmer but with a dimensionless diffusion constant four times larger, $D=2\times 10^{-3}$, and in this case there is but one instance for which the swimmer remains trapped at the surface by the end of the simulation. In the limiting case of very high disorder, diffusive behavior  overwhelms any hydrodynamic effects, and the trajectory essentially behaves as a Brownian motion with reflection on the spherical obstacle.

In Fig.~\ref{TrajectoriesA}a we plot the distance from the surface, $h$, and the pitching angle, $\theta$, as functions of time for two instances in the case $D=2\times 10^{-3}$; we have initialized the system with the body close to the colloid and parallel to the surface, $h(0)=1.001$, and $\theta(0)=0$. In one instance the swimmer stays close to the surface for nearly the duration of the time interval considered while in the second instance the swimmer departs from the surface much earlier. In both cases the distance $h(t)$ does not remain fixed, and instead the body leaves from the spherical surface to distances of variable size repeatedly throughout, though in each case the swimmer is drawn back towards the colloid. The intermittent departures are due to translational fluctuations, and the hydrodynamic attraction rapidly brings the swimmer back to the surface. The rotational diffusion and deterministic dynamics, however, act in concert to rotate the body until it is oriented with nearly the deterministic escape angle, $\theta_{\rm e}=3\alpha/8$ for a spherical swimmer (\S\ref{sec: entrapment and scattering}), at which point a small translational or rotational fluctuation can result in particle escape. We show in Fig.~\ref{TrajectoriesA}b the pitch angle in time for each of the instances shown in Fig.~\ref{TrajectoriesA}a, along with the deterministic escape angle, displayed as a dashed line.

The time spent close to the colloid, or trapping time, is now a random quantity and we seek to understand its distribution. There are at least two natural ways to define trapping times. The first is to measure the first time the swimmer has escaped from the surface out to a specified distance $r$, $T_h=\min_t \{t : h(t)> r\}$, which we refer to as the $h-$trapping time. Alternatively, the trapping time can be studied by looking at the first time that the swimmer reaches a suitable angle for escape in the deterministic setting, $
T_\theta = \min_t \{t : \theta(t)> \theta_{\rm e}\}$, which we refer to as the $\theta$-trapping time. The swimmer may not complete its escape and the dynamics near the wall may include numerous intermittent residences on the surface, a fact that is not captured by this second definition of trapping time. However, $T_\theta$ is easier to analyze than $T_h$, and we have observed in simulations that in many cases the body rotation governs particle escape. In Fig.~\ref{TrajectoriesB} we compare $T_\theta$ to $T_h$ for a threshold value of $r=1.5$ for two cases, $(A,\alpha,D)=(20,0.8,0.002)$ and $(A,\alpha,D)=(80,0.4,0.002)$ (fixing $\alpha^2A$). In the first case, we find that $T_\theta$ is seen to be a nearly perfect proxy for $T_h$ as seen in Fig.~\ref{TrajectoriesA}. For a smaller dipole strength, however, the escape angle is smaller; once the swimmer achieves this orientation it does not swim directly away from the colloid, and instead may reside near the surface for a longer time so that $T_h>T_\theta$. Reducing the threshold value $r$ draws $T_h$ closer to $T_\theta$.

\begin{figure}[ht]
\begin{center}
\includegraphics[width=.44\textwidth]{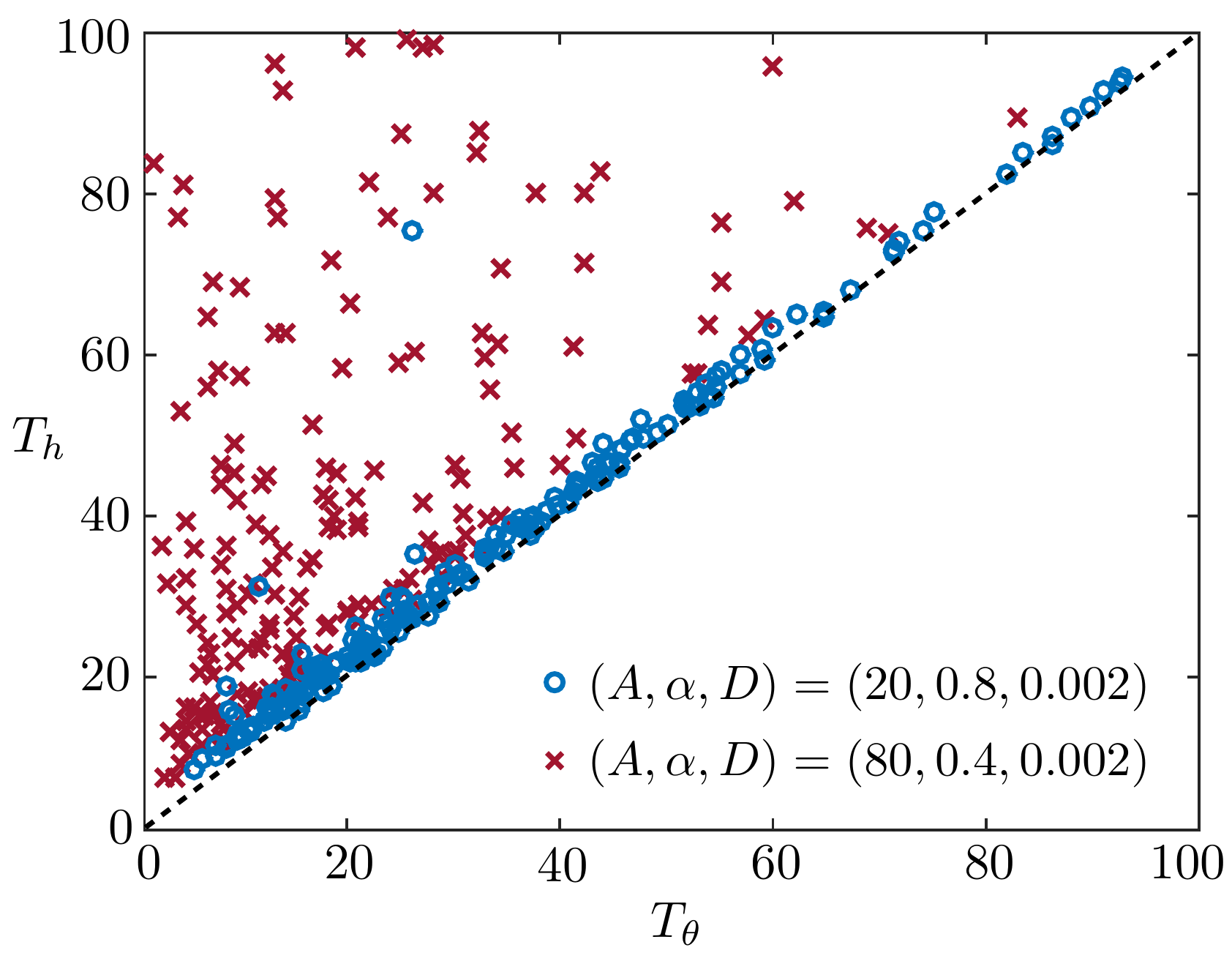}
\caption{Trapping times $T_h$ (with threshold value $r=1.5$) and $T_\theta$ are compared for two cases, from 200 trials. The dashed line indicates $T_h=T_\theta$. A smaller dipole strength corresponds to a smaller escape angle, so that the swimmer resides near the surface for longer before escaping, and $T_h>T_\theta$.}
\label{TrajectoriesB}
\end{center}
\end{figure}

\subsection{Distribution of trapping times}

To gain intuition about the trapping time, we turn to the full simulations. In Fig.~\ref{TrappingTimes} we plot the empirical distributions of the trapping time from $10^4$ independent simulations, where the body is placed initially at $h(0)=1.001$ and parallel with the surface, $\theta(0)=0$. A threshold of $r=1.5$ is chosen in the definition of $T_h$. The distribution depends on the diffusion constant, dipole strength, and colloid size. For $(A,\alpha,D)=(20,0.8,0.002)$ (Fig.~\ref{TrappingTimes}a) it is clear that the distribution is not exponential, which may have been expected, but instead clearly shows a peak at a finite typical escape time. Increasing the diffusion constant to $D=0.004$ (Fig.~\ref{TrappingTimes}b) decreases the expected time, intuitively. The mean escape time is also reduced if instead the dipole strength is reduced ($\alpha=0.2$ in Fig.~\ref{TrappingTimes}c). However, increasing the colloid size to $A=320$ so that $\alpha^2 A$ is identical to that in Fig.~\ref{TrappingTimes}a results in a similar distribution.

In order to understand these empirical distributions, we aim to understand the $\theta$-trapping time, $T_\theta$, by turning to the stochastic differential equation for $\theta$ from Eq.~\eqref{Bswimmer-angle},
\begin{gather}
\frac{d\theta}{dt}= \frac{1}{A}  \cos \theta  - \frac{3\alpha}{16h^3} \sin 2\theta+ \frac{\sqrt{2D}}{A+h}\eta(t) + \sqrt{\frac{3D}{2}}\,\eta_{\rm R}(t),
\end{gather}
where $\eta(t)$ and $\eta_{\rm R}(t)$ are independent one-dimensional Gaussian white noise fluctuations. In the regime $A\gg 1$ the contribution of $\eta(t)$ can be disregarded. Linearizing about $\theta=0$, and setting $h=1$, the pitching angle during contact with the colloid is seen to satisfy an Ornstein-Uhlenbeck process,
\begin{gather}
\frac{d\theta}{dt}=\left(\frac{1}{A}- \frac{3\alpha}{8}\theta\right)+\sqrt{\frac{3D}{2}}\,\eta_{\rm R}(t).\label{dtheta_b2}
\end{gather}

\begin{figure*}[htbp]
\begin{center}
\includegraphics[width=.98\textwidth]{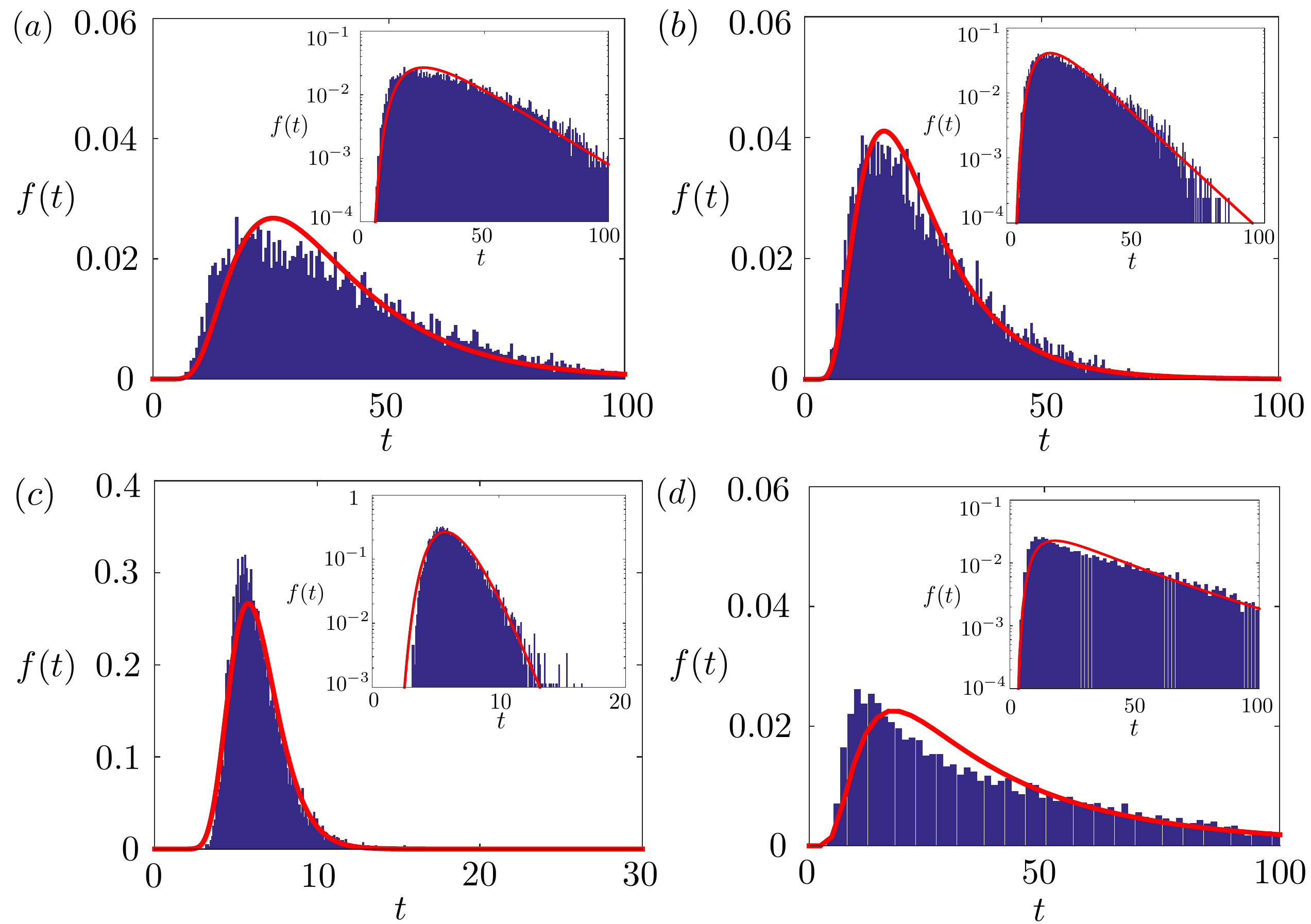}
\caption{Empirical distributions, $f(t)=\partial_t P(T \leq t)$ for the trapping time $T=T_\theta$, from $10^4$ trials by numerical simulation. (a) $(A,\alpha,D)=(20,0.8,0.002)$, with inverse-Gaussian distribution overlaid. The computed mean $\mu$, standard deviation $\sigma$, and shape parameter $\lambda=\mu^3/\sigma^2$ are $(\mu,\sigma,\lambda)=(38.4, 20.3, 137)$. (b) $(A,\alpha,D)=(20,0.8,0.004)$, and inverse-Gaussian distribution with $(\mu,\sigma,\lambda)=(25.1, 13.2, 90.2)$. (c) $(A,\alpha,D)=(20,0.2,0.002)$, with  $(\mu,\sigma,\lambda)=(6.33, 1.61, 98.6)$. (d) $(A,\alpha,D)=(320,0.2,0.002)$, with  $(\mu,\sigma,\lambda)=(43.7, 35.9, 64.7)$. }
\label{TrappingTimes}
\end{center}
\end{figure*}

The distribution of trapping times $f(t)$ (the first passage time) for the Ornstein-Uhlenbeck process with drift, Eq.~\eqref{dtheta_b2}, has been a research topic of its own \cite{Thomas75,nrs85,rs88,app05,dd08,ld08}. There are no known exact expressions for the distribution, with the exception of asymptotically valid distributions and one for a specific parameter relationship.

We draw attention to a few special cases. First, when the diffusion constant $D$ is large, the angle $\theta$ is dominated by the noise term, and the dynamics is primarily governed by a Wiener process. The first passage time of a Wiener process is well studied, it has an inverse-Gaussian distribution, 
\begin{gather}
f(t)=\displaystyle\frac{\lambda}{\sqrt{2\pi t^3}}\,\exp\left(-\frac{\lambda(t-\mu)^2}{2\mu^2 t}\right),
\end{gather}
where $\mu=\mathbb{E}[T]$ is the mean of the distribution and $\lambda=\mu^3/\rm{Var}[T]$ is a shape parameter. For large $D$, $f(t)$ tends towards a L\'evy distribution. A second setting in which the process is approximately governed by a Wiener process is when the colloid size is just larger than the critical size for deterministic entrapment, $A_{\rm c}=64/(9\alpha^2)$. In that case the deterministic component of Eq.~\eqref{dtheta_b2} becomes small and negative as $\theta$ approaches the escape angle, $\theta_{\rm e}=3\alpha/8$. At this point, the determination of the escape time is dominated by diffusion, and we again expect an inverse-Gaussian distribution for the trapping time. In Fig.~\ref{TrappingTimes}a-b we have overlaid on the empirical trapping time distributions the inverse-Gaussian profile, using parameters $\mu$ and $\lambda$ as calculated from the empirical data. Even though the diffusion constant is relatively small, and the colloid size is about twice as large as the critical colloid size, ($A=20$, whereas $A_{\rm c}\approx 11$), the inverse-Gaussian distribution  gives a remarkably accurate depiction of the trapping time in the full simulations. 

A third situation that results in an approximately inverse-Gaussian distribution is when the dipole strength $\alpha>0$ is small, in which case  Eq.~\eqref{dtheta_b2} appears as a Wiener process with drift. Recall that a smaller dipole strength also corresponds to a smaller escape angle. The inverse-Gaussian profile is again seen to match the empirical values closely in Fig.~\ref{TrappingTimes}c, where $\alpha=0.2$. Note that this is not a trapping colloid in the deterministic setting, since $A<A_{\rm c}$, which ensures that the body will escape in finite time even if there are no  fluctuations; this is known as the ``suprathreshhold regime'' \cite{ld08}.

\begin{figure*}[htbp]
\begin{center}
\includegraphics[width=.85\textwidth]{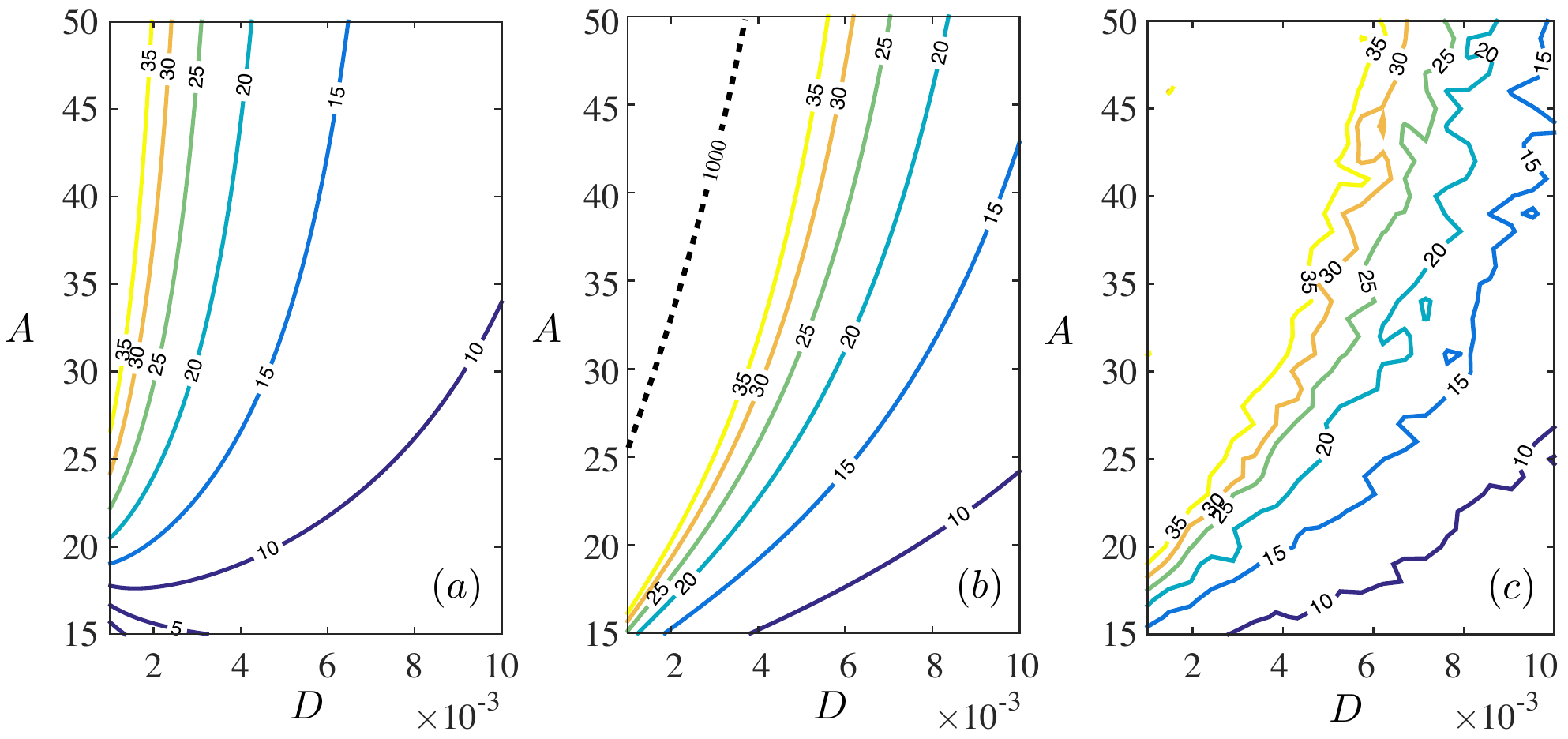}
\caption{Contours of the mean trapping time, $\mathbb{E}[T_\theta]$, with $\alpha=0.8$ as a function of the diffusion constant $D$ and the colloid size $A$, starting from $h(0)=1$ and $\theta(0)=0$: (a) from the simple estimate in Eq.~\eqref{ETtheta}; (b) from numerical integration of Eq.~\eqref{ETtheta_ex}; (c) and from full simulations using 100 trials for each of 720 parameter pairs $(A,D)$ out to a time $t=100$.}
\label{ET_Theta}
\end{center}
\end{figure*}

The small dipole effect can be counteracted, however, by a large colloid size (including the limit of an infinite plane wall). Setting $A=320$ so that $\alpha^2A$ is identical to that used in Fig.~\ref{TrappingTimes}a, the distribution is found to be similar, though with a much longer tail, and the inverse-Gaussian approximation is in fact more accurate here. Had we only focused on Eq.~\eqref{dtheta_b2}, when $A \gg A_c$ and the diffusion constant is not too large, the dynamics are in the ``subthreshhold'' regime and the distribution is well approximated as a Poisson (exponential) distribution \cite{ld08}. The exponential distribution of trapping times was suggested in the model studied by Takagi {\it et al.} \cite{tbzm13}. However, in practice we do not observe an exponential distribution. $T_\theta$ is not a good proxy for $T_h$ when $\alpha$ is relatively small, and Eq.~\eqref{dtheta_b2} does not completely specify the escape dynamics. The issue of escape from an infinite plane wall was also taken up by Drescher {\it et al.} \cite{ddcgg11}, who noted that the escape time is very sensitive to the ratio of translational and rotational diffusion constants, which in turn depend on the distance from the wall. In general, the trapping time distribution from the Ornstein-Uhlenbeck process in Eq.~\eqref{dtheta_b2} resembles something in between exponential and inverse-Gaussian \cite{app05,ld08}.

\subsection{Mean trapping time}

While closed-form expressions of the distribution function are not known for the general case of the Ornstein-Uhlenbeck process, Eq.~\eqref{dtheta_b2}, the moments of the distribution are known\cite{rs88}. It is useful to first linearize the equations around the equilibrium pitching angle on the surface, $\theta^* = 8/(3\alpha A)$, and to define variations around this point as $\tilde{\theta}=\theta-\theta^*$, so that (setting $h=1$),
\begin{gather}
 \frac{d\tilde{\theta}}{dt}= - \frac{3\alpha}{8}\tilde{\theta}+\sqrt{\frac{3D}{2}}\, \eta_{\rm R}(t).\label{dtheta_b3}
\end{gather}
We seek the time for which $\theta=\theta_{\rm e}=3\alpha/8$, the deterministic escape angle (i.e., the first time when $\tilde{\theta}=\tilde{\theta}_e=3\alpha/8-8/(3\alpha A)$). The trapping time $T=T_\theta$ of the Ornstein-Uhlenbeck process with no drift has moments that may be written in a recursive structure in terms of special functions,
\begin{gather}
\b{E}[T^k]=k\int_{\tilde{\theta}_0}^{\tilde{\theta}_e} dz \frac{2}{\sigma^2 W(z)}\int_{-\infty}^z dx W(x)\b{E}[T^{k-1}],\label{ETtheta_ex}
\end{gather}
where
\begin{gather}\label{EqW}
W(x) = \frac{1}{\sigma \sqrt{\pi \tau}}\exp\left(-\frac{x^2}{\sigma^2 \tau}\right),
\end{gather}
and we have defined $\tau=8/(3\alpha)$ and $\sigma=\sqrt{3D/2}$ (see Ref.~\cite{rs88}). An estimate of the mean trapping time may be found by assuming that $\tilde{\theta}_0$ and $\tilde{\theta}_e$ are small. In the event that $\theta(0)=0$, we find
\begin{multline}\label{ETtheta}
\mathbb{E}[T]=\int_{\tilde{\theta}_0}^{\tilde{\theta}_e} dz \frac{2}{\sigma^2 W(z)}\int_{-\infty}^z dx\, W(x)\\
\approx\sqrt{\frac{\pi\alpha}{4D}}+\frac{4}{3AD} \left(\frac{9 \alpha ^2 A}{128}-1\right),
\end{multline}
(see Appendix C). Intuitively, we find that factors which increase the mean trapping time are: smaller diffusion constant, larger dipole strength, and larger colloid size. Yet again, the product $\alpha^2A$ appears; recall the similarity of the distributions in Fig.~\ref{TrappingTimes}a\&d, where $\alpha^2 A$ is fixed.

Figure~\ref{ET_Theta}a shows contours of this simple estimate of the mean trapping time as a function of the diffusion constant and colloid size in the case $\theta(0)=0$. The value computed by integrating Eq.~\eqref{ETtheta} numerically is then displayed in Fig.~\ref{ET_Theta}b, which shows qualitative agreement with the simple estimate, but a considerable departure either when the colloid is large and the diffusion constant is small. Finally, contours of the mean trapping time as determined from simulation of 720 different parameter sets $(A,D)$, each using 100 trials and computing up to $t=100$, are shown in Fig.~\ref{ET_Theta}c, indicating that the linearization of the full system about small $\theta$ used to write Eq.~\eqref{dtheta_b2} gives a very accurate picture of the full dynamics for a wide region of the parameter space.

\section{Conclusion}\label{sec:conc}
In this paper, we have studied the scattering and capture of model 
micro-swimmers by spherical obstacles.  Predictions were given for a critical colloid size, $A_{\rm c}$, as a function of the dipole strength and the body geometry, for which hydrodynamic capture is possible. For situations in which the swimming body is in contact with the colloid but eventually escapes (when $A<A_{\rm c}$), we provided analytical estimates of the residence time near the surface, the escape angle, the distance travelled along the spherical surface, and the net scattering effect of the complete interaction with the colloid. We also investigated the basin of attraction for pushers near the colloid, and while not generally much larger than the spherical radius, we provided a power law scaling of the basin size in terms of the dimensionless parameter $\alpha^2 A$ with exponent $1/5$. The dimensionless number $\alpha^2 A$ featured prominently in our work, including its appearance in the critical colloid size $A_{\rm c}$. Due to the smallness of this attraction region around the sphere for all but the largest colloids and dipole strengths, we expect that entrapment may occur robustly, but only if the particle makes a very direct initial contact with the sphere. This is consistent with the statement by Drescher {\it et al.}~\cite{ddcgg11} that ``hydrodynamics is practically irrelevant if the bacterium is more than a body length away from the surface.'' 

We also considered the contribution of Brownian fluctuations to the dynamics. We demonstrated that a swimmer which would be trapped at the surface in the deterministic case may in the fluctuating case experience an occasional rotation which results in its escape. The residence on the colloid surface can be intermittent, and the colloid may simply act as a pure reflection obstacle in the case of a large diffusion constant. In some cases the residence time was found to be governed by an Ornstein-Uhlenbeck process, which resulted in a trapping time with asymptotic inverse-Gaussian distribution. An analytical estimate of the mean trapping time was derived, comparing favorably to its computed value for a wide range of colloid sizes and diffusion constants. 

\tcb{In addition to Brownian fluctuations, some microorganisms exhibit random changes in their direction at exponentially distributed random times (``run-and-tumble'' locomotion \cite{Berg93}). Geometric defects in synthetic microswimmers can also lead to more complicated random behavior which in turn may have long term consequences for macroscopic diffusion \cite{tbzm13}. The effects of non-Gaussian fluctuations will be considered in future work. In the study of living organisms, flagellar activity may have dramatic effects on entrapment when the body is in contact with a surface \cite{kdpg13}, which presents another interesting direction of study.}

The theory provided in this paper might allow for a more complete model of bacterial populations in an inhomogeneous or porous medium, and we envision applications in bioremediation and microorganism sorting techniques. \tcb{In future experiments, numerous scalings provided in the paper can be tested. Specifically, we hope to see measurements of: the scaling of the critical colloid size for entrapment in the strength of the dipole for both pushers and pullers, the scaling of the basin of attraction with dipole strength and colloid size, the scattering angle as a function of the impact parameter and dipole strength, and the distribution of trapping times in the thermal fluctuations.}

We acknowledge helpful conversations with D.~Takagi, J.~Palacci, 
M.~Shelley, J.~Zhang, and J.-L.~Thiffeault. R.R.~Moreno-Flores acknowledges funding by Fondecyt grant 1130280 and Inicitativa Cientifica Milenio NC130062; and E.~Lauga acknowledges from the European Union through a Marie Curie CIG Grant.

\section{Appendix A: Image system for a no-slip sphere}

The fluid velocity due to a point force of magnitude $\b{f}$ located at at point $\b{y}$ in the fluid, and its image system, derived such that the fluid velocity on the sphere $|\b{x}|=A$ is zero, is written as $\b{u}_j(\b{x})=(\b{S}_{jk}+\b{S^*}_{jk})f_k/(8\pi\mu)$. With $\b{y^*}=(A^2/|\b{y}|^2)\b{y}$ the image point inside the sphere, and $r=|\b{x}-\b{y^*}|$, we have \cite{Oseen27}

\begin{gather}
\b{S}_{jk}=\frac{\delta_{jk}}{|\b{x-y}|}+\frac{(x_j-y_j)(x_k-y_k)}{|\b{x-y}|^3},
\end{gather}
\begin{widetext}
\begin{multline}
\b{S^*}_{jk}f_k=\frac{-A\delta_{jk}}{|\b{y}|r}-\frac{A^3}{|\b{y}|^3}\frac{(x_j-y_j^*)(x_k-y^*_k)}{r^3}-\frac{|\b{y}|^2-A^2}{|\b{y}|}\Big\{\frac{y_j^*y_k^*}{A^3r}-\frac{A}{|\b{y}|^2r^3}[y_j^*(x_k-y_k^*)+y_k^*(x_j-y^*_j)]\\
+\frac{2y^*_j y^*_k y^*_m(x_m-y^*_m)}{A^3 r^3} \Big\}-(|\b{x}|^2-A^2)\Phi,
\end{multline}
\begin{align}
\Phi=&\frac{|\b{y}|^2-A^2}{2|\b{y}|^3}\Big\{-\frac{3(x_j-y^*_j)y_k}{A r^3}+\frac{A\delta_{jk}}{r^3}-3A\frac{(x_j-y^*_j)(x_k-y_k^*)}{r^5}-\frac{2 y^*_j y_k}{A r^3}+\frac{6y_k}{Ar^5}(x_j-y^*_j)y^*_m(x_m-y_m^*)\nonumber\\
&+\frac{3A}{|\b{y}^*|}\frac{(x_j-y^*_j)y_k^* r^2+(x_j-y^*_j)(x_k-y_k^*)|\b{y}^*|^2+(r-|\b{y}^*|)r^2|\b{y}^*|\delta_{jk}}{r^3|\b{y^*}|(r|\b{y^*}|+x_m y_m^*-|\b{y^*}|^2)}\nonumber\\
&-\frac{3A}{|\b{y}^*|}\frac{(|\b{y^*}|(x_j-y^*_j)+r y^*_j)(y_k^*r^2-|\b{y^*}|^2(x_k-y_k^*)+(x_k-2y_k^*)r|\b{y}^*|)}{r^2|\b{y^*}|(r|\b{y^*}|+x_m y^*_m-|\b{y^*}|^2)^2}\nonumber\\
&-\frac{3A}{|\b{y}^*|}\frac{x_j y_k^*+|\b{x}||\b{y^*}|\delta_{jk}}{|\b{x}||\b{y^*}|(|\b{x}||\b{y^*}|+x_m y_m^*)}+\frac{3A}{|\b{y}^*|}\frac{(|\b{y^*}|x_j+|\b{x}|y^*_j)(|\b{y^*}|x_k+|\b{x}|y_k^*)}{|\b{x}||\b{y^*}|(|\b{x}||\b{y^*}|+x_m y_m^*)^2}\Big\}.
\end{align}
\end{widetext}
The velocity field for a symmetric Stresslet and its image system is found by placing two opposing singularities of the form above in the fluid, with strengths inversely proportional to the distance between them, and taking the limit as that distance vanishes.

\section{Appendix B: General expression for translational and angular velocities}

Neglecting the higher order derivatives of the velocity field near the swimming body, we have the following expressions of the hydrodynamic attraction/repulsion and rotation on the swimmer (with $\bm{\tilde{\Omega}}=\tilde{\Omega}\,\b{\hat{r}^\perp}\times \b{\hat{r}}$):

\begin{multline}
\b{\tilde{u}}=\frac{-3 A \alpha(1-3 \sin^2 \theta) (A+h) }{2 h^2 (2 A+h)^2}\b{\hat{r}}\\
+\frac{3 A^3\alpha \left(2 A^2+6 A h+3 h^2\right) \sin 2 \theta}{4 h^2 (A+h)^3 (2 A+h)^2}\b{\hat{r}^\perp},\label{Mainut}
\end{multline}
\begin{multline}
\tilde{\Omega}=\frac{-3\alpha A^3 \sin 2\theta}{4 h^3 (A+h)^2 (2 A+h)^3}\times\\
\left(\left(2 A^2+6 A h+3 h^2\right)+\frac{\Gamma Q(\theta)}{8A^2(A+h)^2} \right),\label{MainOt}
\end{multline}
where
\begin{align}
Q(\theta)=&A^6-5 A^4 (A+h)^2+10 A^2 (A+h)^4\nonumber\\
&+6 (A+h)^6+\Big(9 A^6-29 A^4 (A+h)^2\nonumber\\
&+34 A^2 (A+h)^4-18 (A+h)^6\Big) \cos 2 \theta.
\end{align}

However, if we assume that $A \gg 1$ for fixed $h$ we recover the infinite plane wall result along with the leading order correction for a wall of curvature $1/A$
\begin{multline}
\b{\tilde{u}}=\frac{-3 \alpha  \left(1-\frac{h^2}{4A^2}\right) (1-3 \sin^2 \theta )}{8 h^2}\b{\hat{r}}\\
+\frac{3\alpha \left(1-\frac{h}{A}-\frac{3h^2}{4A^2}\right)\sin 2\theta }{8 h^2}\b{\hat{r}^\perp}+O\left(\frac{\alpha}{h^2}\left(\frac{h}{A}\right)^3\right),\label{simple1}
\end{multline}
\begin{multline}
\tilde{\Omega}=-\frac{3 \alpha \sin 2\theta }{16 h^3}\Bigg\{\left(1-\frac{h}{2A}-\frac{3h^2}{2A^2}\right)\\
-\frac{\Gamma}{2}\left(1+\sin^2 \theta- \frac{h}{A}(1-2 \sin^2 \theta )-\frac{h^2}{A^2}\right) \Bigg\}\\
+O\left(\frac{\alpha}{h^3}\left(\frac{h}{A}\right)^3\right).\label{simple3}
\end{multline}
(See \cite{sl12}). Note that $A\gg 1$ with $h/A$ fixed produces a different expression, but the swimmer may not feel the wall strongly in that case. \\

\section{Appendix C}

The approximating expression for the mean trapping time is found for general initial angle $\theta(0)$ by assuming $\tilde{\theta}_0$ and $\tilde{\theta}_e$ are small, and noting that
\begin{gather}
 \int_{-\infty}^0 W(x) dx =\frac{1}{2},
\end{gather}
for $W(x)$ defined in Eq.~\eqref{EqW}. Taylor expanding about small $z$ in the inner integral of Eq.~\eqref{ETtheta} we have approximately that
\begin{multline}\label{ETtheta_app}
\b{E}[T]=\int_{\tilde{\theta}_0}^{\tilde{\theta}_e} dz \frac{2}{\sigma^2 W(z)}\int_{-\infty}^z dx W(x)\\
\approx \frac{2}{\sigma^2}\int_{\tilde{\theta}_0}^{\tilde{\theta}_e} dz\left\{\frac{1}{2W(0)}+z\left(1-\frac{W'(0)}{2W(0)^2}\right) \right\},
\end{multline}
and then using $W(0)=(\sigma\sqrt{\pi \tau})^{-1}$, $W'(0)=0$, $\tau = 8/(3\alpha)$, and $\sigma =\sqrt{3D/2}$ (and setting $\theta(0)=0$), we arrive at the expression in Eq.~\eqref{ETtheta}.

\bibliographystyle{unsrt}
\bibliography{Bigbib}

\begin{thebibliography}{10}

\bibitem{Rothschild63}
{Ld. Rothschild}.
\newblock Non-random distribution of bull spermatazoa in a drop of sperm
  suspension.
\newblock {\em Nature (London)}, 198:1221--1222, 1963.

\bibitem{fm95}
L.~J. Fauci and A.~McDonald.
\newblock Sperm motility in the presence of boundaries.
\newblock {\em Bull. Math. Biol.}, 57:679--699, 1995.

\bibitem{btbl08}
A.~P. Berke, L.~Turner, H.~C. Berg, and E.~Lauga.
\newblock Hydrodynamic attraction of swimming microorganisms by surfaces.
\newblock {\em Phys. Rev. Lett.}, 101:038102, 2008.

\bibitem{sgbk09}
D.~J. Smith, E.~A. Gaffney, J.~R. Blake, and J.~C. Kirkman-Brown.
\newblock Human sperm accumulation near surfaces: a simulation study.
\newblock {\em J. Fluid Mech.}, 621:289--320, 2009.

\bibitem{sb09}
D.~J. Smith and J.~R. Blake.
\newblock Surface accumulation of spermatozoa: a fluid dynamic phenomenon.
\newblock {\em The Mathematical Scientist}, 34:74--87, 2009.

\bibitem{ddcgg11}
K.~Drescher, J.~Dunkel, L.~H. Cisneros, S.~Ganguly, and R.~E. Goldstein.
\newblock Fluid dynamics and noise in bacterial cell--cell and cell--surface
  scattering.
\newblock {\em Proc. Natl. Acad. Sci. USA}, 108:10940--10945, 2011.

\bibitem{sl12}
S.~E. Spagnolie and E.~Lauga.
\newblock Hydrodynamics of self-propulsion near boundaries: predictions and
  accuracy of far-field approximations.
\newblock {\em J. Fluid. Mech.}, 700:1--43, 2012.

\bibitem{ldlws06}
E.~Lauga, W.~R. DiLuzio, G.~M. Whitesides, and H.~A. Stone.
\newblock Swimming in circles: motion of bacteria near solid boundaries.
\newblock {\em Biophys. J.}, 90:400--412, 2006.

\bibitem{dldaai11}
R.~Di~Leonardo, D.~Dell'Arciprete, L.~Angelani, and V.~Iebba.
\newblock Swimming with an image.
\newblock {\em Phys. Rev. Lett.}, 106:038101, 2011.

\bibitem{gnbnm05}
T.~Goto, K.~Nakata, K.~Baba, M.~Nishimura, and Y.~Magariyama.
\newblock A fluid-dynamic interpretation of the asymmetric motion of singly
  flagellated bacteria swimming close to a boundary.
\newblock {\em Biophys. J.}, 89:3771--3779, 2005.

\bibitem{sgs10}
H.~Shum, E.~A. Gaffney, and D.~J. Smith.
\newblock Modelling bacterial behaviour close to a no-slip plane boundary: the
  influence of bacterial geometry.
\newblock {\em Proc. Roy. Soc. A}, 466:1725--1748, 2010.

\bibitem{giy10}
D.~Giacch{\'e}, T.~Ishikawa, and T.~Yamaguchi.
\newblock Hydrodynamic entrapment of bacteria swimming near a solid surface.
\newblock {\em Phys. Rev. E}, 82:056309, 2010.

\bibitem{znm09}
R.~Zargar, A.~Najafi, and M.~Miri.
\newblock Three-sphere low-{R}eynolds-number swimmer near a wall.
\newblock {\em Phys. Rev. E}, 80(2):026308, 2009.

\bibitem{houg09}
J.~P. Hernandez-Ortiz, P.~T. Underhill, and M.~D. Graham.
\newblock Dynamics of confined suspensions of swimming particles.
\newblock {\em J. Phys.: Condens. Matter}, 21:204107, 2009.

\bibitem{co10}
D.~G. Crowdy and Y.~Or.
\newblock Two-dimensional point singularity model of a low-{R}eynolds-number
  swimmer near a wall.
\newblock {\em Phys. Rev. E}, 81:036313, 2010.

\bibitem{lp10}
I.~Llopis, I. \&~Pagonabarraga.
\newblock Hydrodynamic interactions in squirmer motion: Swimming with a
  neighbour and close to a wall.
\newblock {\em J. Non-Newt. Fluid Mech.}, 165:946--952, 2010.

\bibitem{Crowdy11}
D.~Crowdy.
\newblock Treadmilling swimmers near a no-slip wall at low reynolds number.
\newblock {\em Int. J. Nonlinear Mech.}, 46:577--585, 2011.

\bibitem{vllnz90}
M.~C. Van~Loosdrecht, J.~Lyklema, W.~Norde, and A.~J. Zehnder.
\newblock Influence of interfaces on microbial activity.
\newblock {\em Microbiol. Rev.}, 54:75--87, 1990.

\bibitem{otkk00}
G.~O'Toole, H.~B. Kaplan, and R.~Kolter.
\newblock Biofilm formation as microbial development.
\newblock {\em Annu. Rev. Microbiol.}, 54:49--79, 2000.

\bibitem{hdf92}
G.~Harkes, J.~Dankert, and J.~Feijen.
\newblock Bacterial migration along solid surfaces.
\newblock {\em Appl. Environ. Microbiol.}, 58:1500--1505, 1992.

\bibitem{kdpg13}
V.~Kantsler, J.~Dunkel, M.~Polin, and R.~E. Goldstein.
\newblock Ciliary contact interactions dominate surface scattering of swimming
  eukaryotes.
\newblock {\em Proc. Natl. Acad. Sci. U.S.A.}, 110:1187--1192, 2013.

\bibitem{mbss14}
M.~Molaei, M.~Barry, R.~Stocker, and J.~Sheng.
\newblock Failed escape: Solid surfaces prevent tumbling of escherichia coli.
\newblock {\em Phys. Rev. Lett.}, 113(6):068103, 2014.

\bibitem{gkca07}
P.~Galajda, J.~Keymer, P.~Chaikin, and R.~Austin.
\newblock A wall of funnels concentrates swimming bacteria.
\newblock {\em J. Bacteriol.}, 189:8704--8707, 2007.

\bibitem{wrnr08}
M.~B. Wan, C.~J.~O. Reichhardt, Z.~Nussinov, and C.~Reichhardt.
\newblock Rectification of swimming bacteria and self-driven particle systems
  by arrays of asymmetric barriers.
\newblock {\em Phys. Rev. Lett.}, 101:018102, 2008.

\bibitem{tc09}
J.~Tailleur and M.~E. Cates.
\newblock Sedimentation, trapping, and rectification of dilute bacteria.
\newblock {\em Europhys. Lett.}, 86:60002, 2009.

\bibitem{dladariscmdadf10}
R.~Di~Leonardo, L.~Angelani, D.~DellÕArciprete, G.~Ruocco, V.~Iebba,
  S.~Schippa, M.~P. Conte, F.~Mecarini, F.~De~Angelis, and E.~Di~Fabrizio.
\newblock Bacterial ratchet motors.
\newblock {\em Proc. Natl. Acad. Sci. USA}, 107:9541--9545, 2010.

\bibitem{bjmvdvscm13}
I.~Berdakin, Y.~Jeyaram, V.~V. Moshchalkov, L.~Venken, S.~Dierckx, S.~J.
  Vanderleyden, A.~V. Silhanek, C.~A. Condat, and V.~I. Marconi.
\newblock Influence of swimming strategy on microorganism separation by
  asymmetric obstacles.
\newblock {\em Phys. Rev. E}, 87:052702, 2013.

\bibitem{wlst15}
C.~Wahl, J.~Lukasic, S.E. Spagnolie, and J.-L. Thiffeault.
\newblock Microorganism billiards.
\newblock {\em arXiv preprint arXiv:1502.01478}, 2015.

\bibitem{wwdkg13}
H.~Wioland, F.~G. Woodhouse, J.~Dunkel, J.~O. Kessler, and R.~E. Goldstein.
\newblock Confinement stabilizes a bacterial suspension into a spiral vortex.
\newblock {\em Phys. Rev. Lett.}, 110:268102, 2013.

\bibitem{lwg14}
E.~Lushi, H.~Wioland, and R.~E. Goldstein.
\newblock Fluid flows created by swimming bacteria drive self-organization in
  confined suspensions.
\newblock {\em Proc. Natl. Acad. Sci. USA}, pages 9733--9738, 2014.

\bibitem{pkossacmlc04}
W.~F. Paxton, K.~C. Kistler, C.~C. Olmeda, A.~Sen, S.~K. St.~Angelo, Y.~Cao,
  T.~E. Mallouk, P.~E. Lammert, and V.~H. Crespi.
\newblock Catalytic nanomotors: Autonomous movement of striped nanorods.
\newblock {\em J. Am. Chem. Soc.}, 126(41):13424--13431, 2004.

\bibitem{fbamo05}
S.~Fournier-Bidoz, A.~C. Arsenault, I.~Manners, and G.~A. Ozin.
\newblock Synthetic self-propelled nanorotors.
\newblock {\em Chem. Commun.}, pages 441--443, 2005.

\bibitem{rk07}
G.~R{\"u}ckner and R.~Kapral.
\newblock Chemically powered nanodimers.
\newblock {\em Phys. Rev. Lett.}, 98(15):150603, 2007.

\bibitem{gf09}
A.~Ghosh and P.~Fischer.
\newblock Controlled propulsion of artificial magnetic nanostructured
  propellers.
\newblock {\em Nano Lett.}, 9:2243--2245, 2009.

\bibitem{pgwl11}
O.~S. Pak, W.~Gao, J.~Wang, and E.~Lauga.
\newblock High-speed propulsion of flexible nanowire motors: Theory and
  experiments.
\newblock {\em Soft Matter}, 7:8169--8181, 2011.

\bibitem{dbrfsb05}
R.~Dreyfus, J.~Baudry, M.~L. Roper, M.~Fermigier, H.~A. Stone, and J.~Bibette.
\newblock Microscopic artificial swimmers.
\newblock {\em Nature}, 437:862--865, 2005.

\bibitem{Wang09}
J.~Wang.
\newblock Can man-made nanomachines compete with nature biomotors?
\newblock {\em ACS Nano}, 3:4--9, 2009.

\bibitem{tbzm13}
D.~Takagi, A.~B. Braunschweig, J.~Zhang, and M.~J. Shelley.
\newblock Dispersion of self-propelled rods undergoing fluctuation-driven
  flips.
\newblock {\em Phys. Rev. Lett.}, 110:038301, 2013.

\bibitem{tpbsz13}
D.~Takagi, J.~Palacci, A.~B. Braunschweig, M.~J. Shelley, and J.~Zhang.
\newblock Hydrodynamic capture of microswimmers into sphere-bound orbits.
\newblock {\em Soft Matter}, 10:1784--1789, 2014.

\bibitem{bvdvslp15}
A.~T. Brown, I.~D. Vladescu, A.~Dawson, T.~Vissers, J.~Schwarz-Linek, J.~S.
  Lintuvuori, and W.~C.~K. Poon.
\newblock Swimming in a crystal: Using colloidal crystals to characterise
  micro-swimmers.
\newblock {\em arXiv preprint arXiv:1411.6847}, 2014.

\bibitem{kk91}
S.~Kim and S.~J. Karrila.
\newblock {\em {Microhydrodynamics: Principles and Selected Applications}}.
\newblock Dover Publications, Inc., Mineola, NY, 1991.

\bibitem{ss08}
D.~Saintillan and M.~J. Shelley.
\newblock Instabilities and pattern formation in active particle suspensions:
  Kinetic theory and continuum simulations.
\newblock {\em Phys. Rev. Lett.}, 100:178103, 2008.

\bibitem{lp09}
E.~Lauga and T.R. Powers.
\newblock The hydrodynamics of swimming microorganisms.
\newblock {\em Rep. Prog. Phys.}, 72:096601, 2009.

\bibitem{ss14}
D.~Saintillan and M.~J. Shelley.
\newblock Theory of active suspensions.
\newblock In {\em Complex Fluids in Biological Systems}, pages 319--351.
  Springer, 2015.

\bibitem{bc74}
J.~R. Blake and A.~T. Chwang.
\newblock Fundamental singularities of viscous flow.
\newblock {\em J. Eng. Math}, 8:23--29, 1974.

\bibitem{Blake71}
J.~R. Blake.
\newblock A note on the image system for a stokeslet in a no-slip boundary.
\newblock In {\em Proc. Camb. Phil. Soc}, volume~70, pages 303--310. Cambridge
  Univ Press, 1971.

\bibitem{Oseen27}
C.~W. Oseen.
\newblock {\em Neuere Methoden und Ergbnisse in der Hydrodynamik}.
\newblock Akad.-Verlag, Leipzig, 1927.

\bibitem{ss07}
D.~Saintillan and M.~J. Shelley.
\newblock Orientational order and instabilities in suspensions of
  self-locomoting rods.
\newblock {\em Phys. Rev. Lett.}, 99:058102, 2007.

\bibitem{Thomas75}
M.~U. Thomas.
\newblock Some mean first-passage time approximations for the
  {Ornstein-Uhlenbeck} process.
\newblock {\em J. Appl. Probab.}, pages 600--604, 1975.

\bibitem{nrs85}
A.G. Nobile, L.M. Ricciardi, and L.~Sacerdote.
\newblock Exponential trends of ornstein-uhlenbeck first-passage-time
  densities.
\newblock {\em J. Appl. Probab.}, pages 360--369, 1985.

\bibitem{rs88}
L.~M. Ricciardi and S.~Sato.
\newblock First-passage-time density and moments of the ornstein-uhlenbeck
  process.
\newblock {\em J. Appl. Probab.}, pages 43--57, 1988.

\bibitem{app05}
L.~Alili, P.~Patie, and J.~L. Pedersen.
\newblock Representations of the first hitting time density of an
  ornstein-uhlenbeck process 1.
\newblock {\em Stoch. Model.}, 21:967--980, 2005.

\bibitem{dd08}
S.~Ditlevsen and O.~Ditlevsen.
\newblock {Parameter estimation from observations of first-passage times of the
  Ornstein--Uhlenbeck process and the Feller process}.
\newblock {\em Probabilist. Eng. Mech.}, 23:170--179, 2008.

\bibitem{ld08}
P.~Lansky and S.~Ditlevsen.
\newblock A review of the methods for signal estimation in stochastic diffusion
  leaky integrate-and-fire neuronal models.
\newblock {\em Biol. Cybern.}, 99:253--262, 2008.

\bibitem{Berg93}
H.~C. Berg.
\newblock {\em {Random Walks in Biology}}.
\newblock Princeton University Press, 1993.

\end{thebibliography}
\end{document}